\documentclass[prd,twocolumn,showpacs,preprintnumbers,nofootinbib,superscriptaddress,amsmath,amssymb,
aps]{revtex4} 
\usepackage{mathrsfs}
\usepackage{graphicx}
\usepackage{dcolumn}
\usepackage{bm}
\renewcommand{\labelenumi}{(\roman{enumi})}
\usepackage{comment}
\usepackage{braket}
\usepackage{slashed}

\begin{document}

\preprint{CHIBA-EP-245, 2020.05.27}

\title{Complex poles and spectral functions of Landau gauge QCD and QCD-like theories
}

\author{Yui Hayashi}
\email{yhayashi@chiba-u.jp}
\affiliation{
Department of Physics, Graduate School of Science and Engineering, Chiba University, Chiba 263-8522, Japan
}

\author{Kei-Ichi Kondo}
\email{kondok@faculty.chiba-u.jp}
\affiliation{
Department of Physics, Graduate School of Science and Engineering, Chiba University, Chiba 263-8522, Japan
}
\affiliation{
Department of Physics, Graduate School of Science, Chiba University, Chiba 263-8522, Japan
}


\begin{abstract}
In view of the expectation that the existence of complex poles is a signal of confinement, we investigate the analytic structure of the gluon, quark, and ghost propagators in the Landau gauge QCD and QCD-like theories by employing an effective model with a gluon mass term of the Yang-Mills theory, which we call the massive Yang-Mills model. 
In this model, we particularly investigate the number of complex poles in the parameter space of the model consisting of gauge coupling constant, gluon mass, and quark mass for the gauge group $SU(3)$ and various numbers of quark flavors $N_F$ within the asymptotic free region.
Both the gluon and quark propagators at the best-fit parameters for $N_F=2$ QCD have one pair of complex conjugate poles, while the number of complex poles in the gluon propagator varies between zero and four depending on the number of quark flavors and quark mass.
Moreover, as a general feature, we argue that the gluon spectral function of this model with nonzero quark mass is negative in the infrared limit.
In sharp contrast to gluons, the quark and ghost propagators are insensitive to the number of quark flavors within the current approximations adopted in this paper. 
These results suggest that details of the confinement mechanism may depend on the number of quark flavors and quark mass.
\end{abstract}

\maketitle


\section{INTRODUCTION}
Color confinement, absence of color degrees of freedom from the physical spectrum, is one of the most fundamental and significant features of strong interactions. It is a long-standing and challenging problem in particle and nuclear physics to explain color confinement in the framework of quantum field theory (QFT). Understanding analytic structures of the correlation functions will be of crucial importance to this end because a QFT describing physical particles can be reformulated in terms of correlation functions \cite{AQFT}, and there are some proposals of confinement mechanisms whose criteria are expressed by them, e.g., \cite{KO79}. In particular, the analytic structures of propagators encode kinematic information as the K\"all\'en-Lehmann spectral representation \cite{spectral_repr_UKKL}, which will be useful toward understanding confinement.

In the past decades, numerous studies of both the lattice and continuum approaches have focused on the gluon, quark, and ghost propagators in the Landau gauge of the Yang-Mills theory and quantum chromodynamics (QCD). In the Yang-Mills theory, or the quenched limit of QCD, the so-called \textit{decoupling} and \textit{scaling} solutions of the gluon and ghost propagators are observed based on the continuum approaches \cite{decoupling-analytical}. The recent numerical lattice results support the decoupling solution \cite{decoupling-lattice}.
The decoupling solution has an impressing feature that the running gauge coupling stays finite and nonzero for all nonvanishing momenta and eventually goes to zero in the limit of vanishing momentum, which cannot be predicted from the standard perturbation theory that is plagued by the Landau pole of the diverging running gauge coupling.

A low-energy effective model of the Yang-Mills theory is proposed following the decoupling behavior of the gluon propagator, which provides the gluon and ghost propagators that show a striking agreement with the numerical lattice results by including quantum corrections just in the one-loop level \cite{TW10,TW11}. This effective model is given by the mass-deformed Faddeev-Popov Langrangian of the Yang-Mills theory in the Landau gauge, or the Landau gauge limit of the Curci-Ferarri model \cite{CF76b}, which can be shown to be renormalizable due to the modified Becchi-Rouet-Stora-Tyutin (BRST) symmetry, and we call it the \textit{massive Yang-Mills model} for short. This effective mass term could stem from the dimension-two gluon condensate \cite{Schaden-Kondo-Shinohara-Wershinke,Boucaud2000,Gubarev-Zakharov,Verschelde2001,BG2003} or could be taken as a (minimal) consequence of avoiding the Gribov ambiguity \cite{Gribov78,ST12}.
Moreover, it has been shown that the massive Yang-Mills model has ``infrared safe'' renormalization group (RG) flows on which the running gauge coupling remains finite for all scales \cite{TW11,RSTW17,ST12}. The three-point functions \cite{PTW13} and two-point correlation functions at finite temperature \cite{RSTW14} in this model were compared to the numerical lattice results, showing good agreements. Moreover, the two-loop corrections improve the accordance for the gluon and ghost propagators \cite{GPRT19}. These works indicate the validity of the massive Yang-Mills model as an effective model of the Yang-Mills theory.

The gluon and ghost propagators for unquenched lattice QCD with the number of quark flavors $N_F = 2,~ 2+1,~2+1+1$ have been studied, for instance \cite{unquenched-lattice}, and exhibit the decoupling feature as well. The massive Yang-Mills model with dynamical quarks can reproduce the numerical lattice gluon and ghost propagators for QCD as well \cite{PTW14}. However, it is argued that higher-loop corrections are important for the quark sector in this model \cite{PTW15, PRSTW17}. Despite this shortcoming, it appears that the effective gluon mass term captures some nonperturbative aspects of QCD. What is more, QCD phases have been extensively studied in a similar model with the effective gluon mass term of the Landau-DeWitt gauge~\cite{Finite-temperature-Landau-dewitt}.

Apart from the realistic QCD, it is also interesting to study gauge theories with many flavors of quarks. For many quark flavors, the infrared conformality is predicted \cite{BZ82,Caswell} and well-studied in line with the walking technicolor for physics beyond the standard model \cite{Technicolor}, for example, \cite{conformal-window-lat,conformal-window-sd,conformal-window-frg}. Some argue that chiral symmetry restores while color degrees of freedom are ``unconfined'' in the conformal window. For a better understanding of the confinement mechanism, observing $N_F$ dependence will be thus extremely valuable.

All works on the correlation functions described above were implemented in the Euclidean space. Considerable efforts have been devoted to reconstructing the spectral functions from the Euclidean data based on the K\"all\'en-Lehmann spectral representation, e.g., \cite{spectral-numerical,CPRW18}.
On the other hand, several models of the Yang-Mills theory \cite{Gribov78,Zwanziger90,DGSVV2008,Siringo16a,Siringo16b,Stingl86,Stingl96,HKRSW}, including the (pure) massive Yang-Mills model \cite{HK2018,KSOMH18}, and a way of the reconstruction from the Euclidean data \cite{BT2019} predict complex poles in the gluon propagator that invalidate the K\"all\'en-Lehmann spectral representation.
The existence of complex poles of the propagators of the confined particles is a controversial issue, see, e.g., \cite{SFK12}.
The complex singularities invalidate the standard reconstruction from a Euclidean field theory to a relativistic QFT \cite{AQFT} and might correspond to unphysical degrees of freedom in an indefinite metric state space \cite{Nakanishi72suppl}. Therefore, complex poles are expected to be closely connected to the confinement mechanism.

In this paper, we investigate the analytic structure of the QCD propagators for various $N_F$ based on the massive Yang-Mills model, mainly focusing on complex poles by utilizing the general relationship between the number of complex poles and the propagator on timelike momenta from the argument principle \cite{HK2018}. This investigation extends the previous result \cite{HK2018} obtained for the pure Yang-Mills theory with no flavor of quarks $N_F=0$ that the gluon propagator has a pair of complex conjugate poles and the negative spectral function while the ghost propagator has no complex pole. In this article, we will see the following results.
Both the gluon and quark propagators at the ``realistic'' parameters for $N_F=2$ QCD have one pair of complex conjugate poles as well as the gluon propagator in the zero flavor case.
By increasing quark flavors, we find a new region in which the gluon propagator has two pairs of complex conjugate poles for light quarks with the intermediate number of flavors $4 \lesssim N_F<10$. 
However, the gluon propagator has no complex poles if very light quarks have many flavors $N_F \geq 10$ or both of the gauge coupling and quark mass are small. 
In the other regions, the gluon propagator has one pair of complex conjugate poles. On the other hand, the analytic structures of quark and ghost propagators are nearly independent of the number of quarks within this one-loop analysis.

This paper is organized as follows. In Sec.~II, we present machinery to count the number of complex poles as an application of \cite{HK2018}. In Sec.~III, we review the calculation of the massive Yang-Mills model with quarks of \cite{PTW14}, consider the infrared safe trajectories of this model, and argue the infrared negativity of the gluon spectral function. 
Then, in Sec.~IV, we analyze the analytic structures of the gluon, quark, and ghost propagators at the best-fit parameters of \cite{PTW14}, and investigate the number of complex poles for various parameters and the number of quarks $N_F$. Finally, Sec.~V is devoted to conclusion, and Sec.~VI contains further discussion. In Appendix A, we provide a generalization of the proposition of Sec.~II for various infrared behaviors. Appendix B gives complementary one-loop analyses for Sec.~IV A.

\section{COMPLEX POLES IN PROPAGATORS}
To elucidate the starting point, we review a generalization of the spectral representation for a propagator so as to allow complex poles. We then develop a method for counting the number of complex poles from the data on timelike momenta, e.g., the spectral function, as a straightforward application of the general relation \cite{HK2018}.

\subsection{A generalization of the spectral representation}

We introduce some definitions and underlying assumptions on propagators adopted in this article. Given a propagator defined in the Euclidean space, we analytically continue the propagator to the whole complex $k^2$ plane from the Euclidean momenta. In the complex $k^2$ plane, we call points on \textit{the negative real axis} \textit{Euclidean momenta} and points on \textit{the positive real axis} \textit{timelike momenta}. 
We will study the gluon, quark, and ghost propagators in the Landau gauge. We assume each propagator of the gluon, quark, and ghost has the following generalized spectral representation allowing the presence of complex poles:\footnote{This generalization can be related to the fact that complex spectra for confined particles need not be excluded in an indefinite metric state space \cite{Nakanishi72suppl}.
Such complex spectra of a Hamiltonian can give rise to the complex poles.
The kinematic aspects of complex poles will be discussed elsewhere. Incidentally, another generalization for the QCD propagators within the framework of tempered distribution is proposed in \cite{Lowdon}.}
\begin{align}
 D(k^2) &= \int_0 ^\infty d \sigma^2 \frac{\rho(\sigma^2)}{\sigma^2 - k^2} + \sum_{\ell=1}^n \frac{Z_\ell}{z_\ell - k^2}, \label{eq:generalized-spec-repr}
\end{align}
where $\rho(\sigma^2) := \frac{1}{\pi} \operatorname{Im} D(\sigma^2 + i \epsilon)$ is its spectral function, $z_\ell$ stands for the position of a complex pole, and  $Z_\ell$ is the residue associated with the complex pole.

\begin{figure}[tbp]
 \begin{center}
 \includegraphics[width=0.9\linewidth]{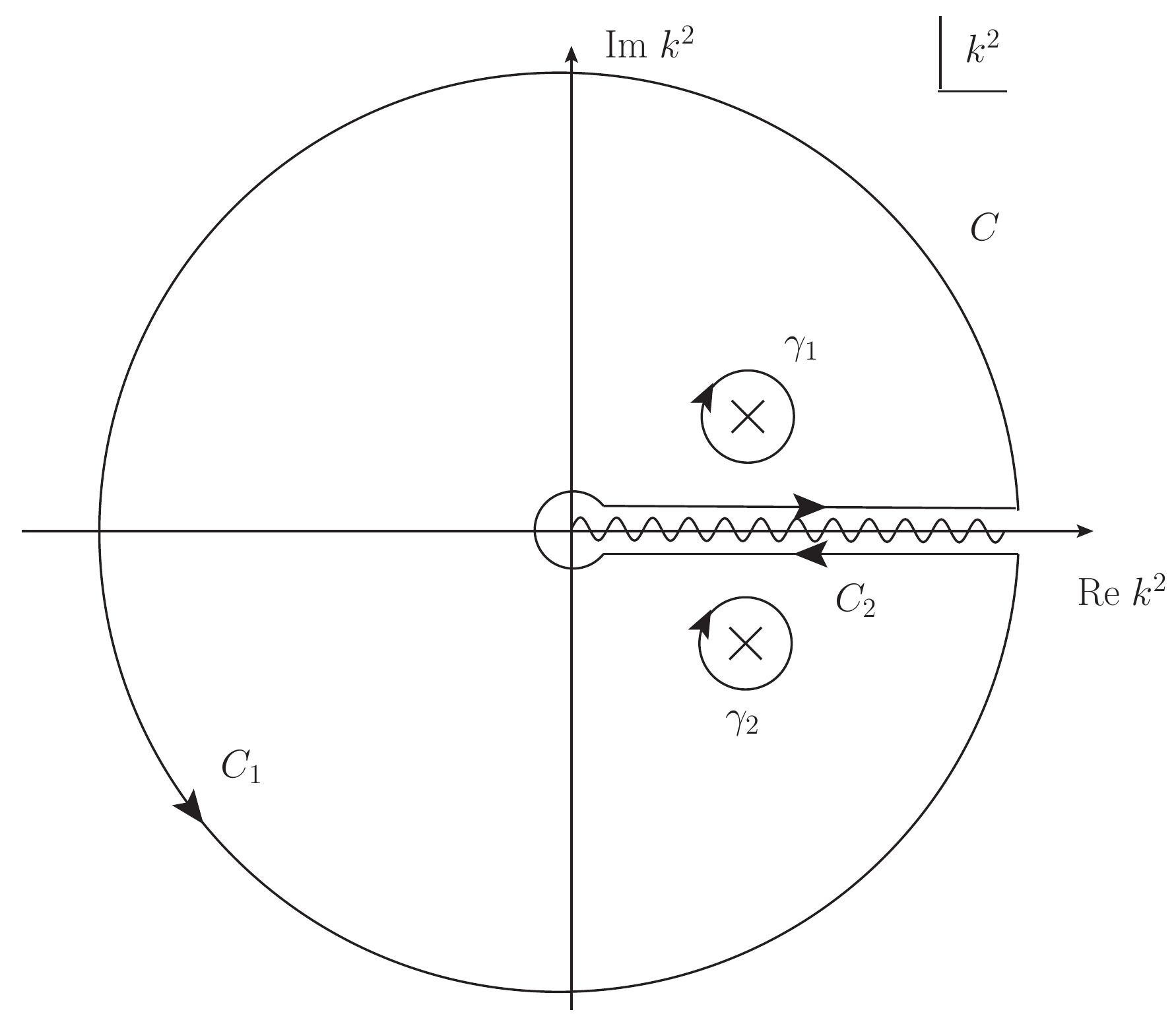}
 \end{center}
 \caption{
 Contour $\tilde{C}$ on the complex $k^2$ plane avoiding the singularities on the positive real axis and the complex poles. The contour $\tilde{C}$ consists of the large circle $C_1$, the path wrapping around timelike singularities $C_2$, and the small contours $\gamma_\ell $ that clockwise surround the complex poles at $z_\ell$. The propagator $D(k^2)$ is holomorphic in the region bounded by the contour $\tilde{C} = C \cup \{ \gamma_\ell \}_{\ell = 1} ^n$, where we denote the closed contour $C_1 \cup C_2$ by $C$.}
 \label{fig:section2_complex.eps}
\end{figure}

This representation can be derived under the following conditions using the Cauchy integral formula for the closed contour $\tilde{C}$ presented in Fig.~\ref{fig:section2_complex.eps}  \cite{HK2018, Siringo17a}:

\begin{enumerate}
 \renewcommand{\labelenumi}{(\Roman{enumi})}
 \item $D(z)$ is holomorphic except for singularities on the positive real axis and a finite number of simple poles.
 \item $D(z) \rightarrow 0$ as $|z| \rightarrow \infty$.
 \item $D(z)$ is real on the negative real axis.
\end{enumerate}
If we replace the condition (I) with the more strict condition 
\begin{enumerate}
 \renewcommand{\labelenumi}{(\Roman{enumi}')}
    \item $D(z)$ is holomorphic except for singularities on the positive real axis, namely has no complex poles,
\end{enumerate}
the three conditions (I'), (II), and (III) lead to the K\"all\'en-Lehmann form~\cite{spectral_repr_UKKL}, which propagators for unconfined particles are supposed to obey. In this sense, eq.~(\ref{eq:generalized-spec-repr}) gives a generalization of the K\"all\'en-Lehmann spectral representation.

\subsection{Counting complex poles}
We present a procedure to count complex poles from the propagator on the timelike momenta based on \cite{HK2018}.

We apply the argument principle to a propagator on the contour $C$ presented in Fig.~\ref{fig:section2_complex.eps}. Then the {\it winding number} $N_W(C)$ of the phase of the propagator $D(k^2)$ along the contour $C$ is equal to the difference between the number of zeros $N_Z$ and the number of poles $N_P$ in the region bounded by $C$,
\begin{align}
N_W(C) :&= \frac{1}{2 \pi i} \oint_{C} d k^2 \frac{D'(k^2)}{D(k^2)} \notag \\
&= \frac{1}{2 \pi} \oint_{C} d (\arg D(k^2))\notag \\
&= N_Z - N_P. \label{eq:winding_number_relation}
\end{align}
The winding number $N_W(C)$ can be calculated from the propagator on timelike momenta $D(k^2 + i \epsilon)$ and infrared (IR) and ultraviolet (UV) asymptotic forms.

In what follows, we assume the following asymptotic form for the propagator.
\begin{enumerate}
 \item \footnote{This assumption (i) is the same as assumption (i) of the assertions in Sec. III in \cite{HK2018}.} In the UV limit $|k^2| \rightarrow \infty$, $D(k^2)$ has the same phase as the free one, i.e., $\arg (-D(z)) \rightarrow  \arg \frac{1}{z} $ as $|z| \rightarrow \infty$.
 \item In the IR limit $|k^2| \rightarrow 0$,  $D(k^2 = 0) > 0$.
\end{enumerate}
Let us comment on these assumptions. The first assumption is satisfied by the gluon, quark, and ghost propagators in the Landau gauge, which follows from the RG analysis for asymptotic free theories. The RG argument of Oehme and Zimmermann \cite{OZ80} provides the following UV asymptotic form for the propagators, as $|k^2| \rightarrow \infty$,
\begin{align}
D(k^2) \simeq - \frac{Z_{UV}}{k^2 (\ln |k^2|)^\gamma}, \label{eq:section2_UV_asymptotic}
\end{align}
where $Z_{UV}$ is a positive constant and $\gamma = \gamma_0/\beta_0$ is the ratio between the first coefficients $\gamma_0, \beta_0$ of the anomalous dimension and the beta function, respectively. For the gluon propagator of Yang-Mills theories with $N_F$ quarks in the Landau gauge, $\gamma$ is computed as follows.
\begin{align}
\gamma &= \frac{\gamma_0}{\beta_0}, \notag \\
\gamma_0 &= - \frac{1}{16 \pi^2} \left( \frac{13}{6} C_2 (G) - \frac{4}{3} C(r) \right),\notag \\
\beta_0 &= - \frac{1}{16 \pi^2} \left( \frac{11}{3} C_2 (G) - \frac{4}{3} C(r) \right),
 \label{eq:gamma_0_and_beta_0}
\end{align}
where $C_2(G)$ and $C(r) = N_F/2$ are the Casimir invariants of the adjoint and fundamental representations of the gauge group $G$, respectively.

We restrict ourselves to asymptotically free theories: $\beta_0 < 0$, or $N_F < \frac{11}{2} C_2(G)$, which is essential to derive the above UV asymptotic expression for the propagator (\ref{eq:section2_UV_asymptotic}).

Note that the sign of $\gamma$ is determined depending on the number of quarks as follows.
\begin{itemize}
    \item $\gamma > 0$ if $C(r) < \frac{13}{8} C_2(G)$. For the gauge group $SU(3)$ with $C_2(G) = 3$, in particular, $\gamma > 0$ if $N_F = 1, \cdots , 9 < \frac{13}{4} C_2(G) = \frac{39}{4}$.
    \item $\gamma < 0$ if $\frac{13}{8} C_2(G) < C(r) < \frac{11}{4} C_2(G)$. For the gauge group $SU(3)$ with $C_2(G) = 3$, in particular, $\gamma < 0$ if $N_F = 10, \cdots , 16 < \frac{11}{2} C_2(G) = \frac{33}{2}$.
\end{itemize}
This determines the sign of the gluon spectral function $\rho(\sigma^2)$ in the ultraviolet region~\cite{OZ80}. The spectral function of the gluon for $SU(3)$ in the UV limit takes negative values for $N_F = 1, \cdots , 9$ and positive values for $N_F = 10, \cdots , 16$.

Additionally, the one-loop corrections give no contribution to the quark anomalous dimension $\gamma_\psi$ in the Landau gauge: $\gamma_\psi = O(g^4)$. Therefore, the quark propagator behaves in the UV limit as the free one because of the asymptotic freedom and the vanishing of the first coefficient $\gamma_\psi^0$ of the order $g^2$ of the quark anomalous dimension.

The second assumption (ii) indicates the massive-like behavior for the propagator, which corresponds to the decoupling solution for the Euclidean gluon propagator. For the general cases $D(k^2) \rightarrow Z_{IR} (-k^2)^\alpha$ as $|k^2| \rightarrow 0$ with a real exponent $\alpha$, e.g., the scaling solution for the pure Yang-Mills gluon and a massless propagator, there is an additional contribution to the expression of $N_W(C)$ described below (\ref{eq:winding-timelike}). See Appendix A for the details of the additional term. From here on, we simply assume (ii) and will verify this assumption when we compute $N_W(C)$ for each propagator employed.

Let us add notes on the relationships between the number $N_W(C)$ and the spectral function \cite{HK2018}.
With the assumptions (i) and (ii), the positive spectral function implies $N_W(C) = 0$ and the negative one implies $N_W(C) = -2$. Since the winding number is a topological invariant, $N_W(C)$ is invariant under continuous deformations. For example, if the spectral function is ``quasinegative'', i.e., the spectral function is negative $\rho(k_0^2) < 0$ at all real and positive zeros $k_0^2$ of ${\rm Re}\ D(k^2)$ i.e., ${\rm Re}\ D(k_0^2) = 0$ ($k_0^2 > 0$), then the propagator has $N_W(C) = -2$. Actually, this is the case of the massive Yang-Mills model with $N_F = 2$ quarks at the realistic parameters analyzed below.

In order to calculate the winding number $N_W(C)$ in a numerical way, we divide the interval $[\delta^2, \Lambda^2]$ on the positive real axis into $(N+1)$ segments $x_0,x_1,\cdots,x_{N+1}$ such that the following condition on $\{ D(x_n + i \epsilon) \}_{n=1}^N$ at points $\{k^2 = x_n + i \epsilon \}_{n=1}^N$ is satisfied.
\begin{enumerate}
\setcounter{enumi}{2}
 \item $\{k^2 = x_n + i \epsilon \}_{n=0}^N$ is sufficiently dense so that $D(k^2= x + i \epsilon)$ changes its phase at most half-winding ($\pm \pi$) between $x_n+i\epsilon$ and $x_{n+1}+i\epsilon$, i.e., for $n = 0, 1, \cdots, N$,
 \begin{align}
\left| \int_{x_n}^{x_{n+1}} dx \frac{d}{dx} \arg D(x + i \epsilon) \right| < \pi,
\end{align}
where we denote sufficiently small $x_0 = \delta^2 >0$ and sufficiently large $x_{N+1} = \Lambda^2$, on which we will take the limits $\delta^2 \rightarrow +0$ and $\Lambda^2 \rightarrow +\infty$.
\end{enumerate}

Let us now calculate $N_W(C)$ from the data $\{ D(x_n + i \epsilon) \}_{n=1}^N$ under the above assumptions, by evaluating $N_W(C_1)$ and $N_W(C_2)$ separately, where $C_1$ stands for the large circle and $C_2$ for the path around the positive real axis depicted in Fig. \ref{fig:section2_complex.eps}.

The first assumption (i) yields
 \begin{align}
N_W (C_1) = -1.
\end{align}

For $N_W(C_2)$, from the Schwarz reflection principle, $[D(z)]^* = D(z^*)$, we have,
 \begin{align}
N_W (C_2) = 2 \int_0^\infty \frac{dx}{2 \pi} \frac{d}{dx} \arg D(x + i \epsilon). \label{eq:sect2_nec2}
\end{align}
Notice that we have used the second assumption (ii) to eliminate the contribution from the small circle around the origin.
The third assumption (iii) transforms the integral (\ref{eq:sect2_nec2}) into a discrete sum as
\begin{align}
N_W (C_2) = 2 \sum_{n=0}^N \frac{1}{2\pi} \operatorname{Arg}\left[ \frac{D(x_{n+1}+i\epsilon)}{D(x_n + i \epsilon)}\right],
\end{align}
where $\operatorname{Arg}$ is the principal value of the argument ($-\pi < \operatorname{Arg} z <\pi$).

To sum up, we have the expression for $N_W(C)$,
\begin{align}
N_W (C) = -1 + 2 \sum_{n=0}^N \frac{1}{2\pi} \operatorname{Arg}\left[ \frac{D(x_{n+1}+i\epsilon)}{D(x_n + i \epsilon)}\right]. \label{eq:winding-timelike}
\end{align}
Though we have to know the number of complex zeros for counting the exact number of complex poles $N_P$ from the winding number $N_W(C)$, note that it suffices to verify $N_W(C) < 0$ in order to show the existence of complex poles because $N_P = N_Z - N_W(C) \geq - N_W(C)$.
Moreover, since $N_W(C)$ is invariant under continuous deformations, we can expect that $N_W(C)$ should be robust under some approximations. We will consider complex poles of the QCD propagators using an effective model and the relation (\ref{eq:winding-timelike}) based on this expectation.

\section{MASSIVE YANG-MILLS MODEL AS AN EFFECTIVE MODEL}
The massive Yang-Mills model \cite{TW10,TW11} to be defined shortly is an effective model of the Yang-Mills theory in the Landau gauge, which captures some nonperturbative aspects by introducing a phenomenological mass term for the gluon.
The mass term may be generated by the effect of the dimension-two gluon condensate \cite{Schaden-Kondo-Shinohara-Wershinke,Boucaud2000,Gubarev-Zakharov,Verschelde2001,BG2003} or by avoiding the Gribov ambiguity \cite{ST12}.
For the latter effect, intuitively, some effect suppressing $\int d^D x ~ {\mathscr A}^A_\mu {\mathscr A}^A_\mu $ should be taken into account due to the Gribov copies, and the most infrared relevant term of this effect will be the mass term. 
The massive Yang-Mills model reproduces the numerical lattice results of the gluon and ghost propagators well and has a renormalization condition such that there are RG trajectories whose running gauge coupling remains finite in all scales. Moreover, the massive Yang-Mills model gives a correct UV asymptotic behavior \cite{OZ80} by RG improvement, as we will see in Sec.~III C.
Therefore, although we expect the massive Yang-Mills model is a low-energy effective model, this model can describe Yang-Mills theory in all scales to some extent. 
One might worry about the absence of the nilpotent BRST symmetry. Although the massive Yang-Mills model suffers from the problem of physical unitarity \cite{CF76b, Kondo13} as a consistent QFT, we insist that this model will suffice for our purpose of investigating analytic structures, as it gives the well-approximating propagators with a sensible and straightforward mass-deformation.

Moreover, by adding effective quark mass term, this model has good accordance with the numerical lattice results for the unquenched gluon and ghost propagators and can also reproduce the quark mass function qualitatively \cite{PTW14}. In this section, we review the results on the one-loop computations of the massive Yang-Mills model with quarks, prepare expressions that we will use in the investigation of the analytic structures of the propagators in the next section, and study asymptotic behaviors of the propagators of this model using RG.

In the Euclidean space, the Lagrangian of the model is given by \cite{TW10,TW11,PTW14}
\begin{align}
{\mathscr L}_{mYM} &= {\mathscr L}_{YM} + {\mathscr L}_{GF} + {\mathscr L}_{FP} + {\mathscr L}_{m}+ {\mathscr L}_{q}, \\
{\mathscr L}_{YM} &= \frac{1}{4} {\mathscr F}^A_{\mu \nu} {\mathscr F}^{A}_{\mu \nu}, \notag \\
{\mathscr L}_{GF} &= i \mathscr{N}^A \partial_\mu {\mathscr A}^A_\mu \notag \\
{\mathscr L}_{FP} &=  \bar{{\mathscr C}}^A \partial_\mu {\mathscr D}_\mu[{\mathscr A}]^{AB} {\mathscr C}^B \notag \notag \\
&=  \bar{{\mathscr C}}^A \partial_\mu (\partial_\mu {\mathscr C}^A + g_b f^{ABC} {\mathscr A}^B_\mu {\mathscr C}^C) \notag \\
{\mathscr L}_{m} &= \frac{1}{2} M^2_b {\mathscr A}_\mu^A {\mathscr A}_\mu^A, \notag  \\
{\mathscr L}_{q} &= \sum_{i=1}^{N_F} \bar{\psi}_i (\gamma_\mu {\mathscr D}_\mu[{\mathscr A}] + (m_b)_{q,i}) \psi_i \notag \\
&= \sum_{i=1}^{N_F} \bar{\psi}_i (\gamma_\mu (\partial_\mu - i g_b {\mathscr A}_\mu^A t^A) + (m_b)_{q,i}) \psi_i,
\end{align}
where we have introduced the bare gluon, ghost, anti-ghost, Nakanishi-Lautrup, and quark fields denoted by ${\mathscr A}^A_\mu, ~ {\mathscr C}^A , ~ \bar{{\mathscr C}}^A, \mathscr{N}^A, \psi_i$ respectively, the bare gauge coupling constant $g_b$, the bare quark mass $(m_b)_{q,i}$, and the bare effective gluon mass $M_b$, while $f^{ABC}$ stands for the structure constant associated with the generators $t^A$ of the fundamental representation of the group $G$.

We introduce the renormalization factors $(Z_A,Z_C, Z_{\bar{C}} = Z_C, Z_{\psi}^{(i)}),~ Z_g,~ Z_{M^2}, ~ Z_{m_{q,i}}$ for the gluon, ghost, anti-ghost, and quark fields $({\mathscr A}_\mu, {\mathscr C}, \bar{{\mathscr C}}, \psi_i)$, the gauge coupling constant $g$, and the gluon and quark mass parameters $M^2, m_{q,i}$ respectively:
\begin{align}
{\mathscr A}^\mu &= \sqrt{Z_A} {\mathscr A}_R^\mu, ~ {\mathscr C} = \sqrt{Z_C} {\mathscr C}_R, \notag \\
\bar{{\mathscr C}} &= \sqrt{Z_C} \bar{{\mathscr C}}_R, ~\psi_i =  \sqrt{Z_{\psi}^{(i)}}\psi_{R,i}, \notag \\
~g_b &= Z_g g, ~M^2_b = Z_{M^2} M^2, ~ (m_b)_{q,i} = Z_{m_{q,i}} m_{q,i}
\end{align}
Throughout this article, for simplicity, we employ this model with degenerate quark masses, $m_q := m_{q,i}$, and therefore $Z_{\psi} := Z_{\psi}^{(i)}$ and $Z_{m_{q}} := Z_{m_{q,i}}$.

\subsection{Strict one-loop calculations}
We review the strict one-loop results for the gluon, quark, and ghost propagators here and the RG functions in the next subsection \cite{PTW14}.

For the gluon, ghost, and quark, we introduce the two-point vertex functions $\Gamma_{{\mathscr A}}^{(2)}$, $\Gamma_{gh}^{(2)}$, and $\Gamma_{\psi}^{(2)}$, the transverse gluon propagator ${\mathscr D}_T$, the ghost propagator $\Delta_{gh}$, the quark propagator $\mathcal{S}$, dimensionless gluon and ghost vacuum polarizations $\hat{\Pi}$ and $\hat{\Pi}_{gh}$, and the scalar and vector part of the quark two-point vertex function $\Gamma_{s}^{(2)},~\Gamma_{v}^{(2)}$ as 
\begin{align}
\Gamma_{{\mathscr A}}^{(2)}(k_E^2) &:= [{\mathscr D}_T (k_E^2)]^{-1} \notag \\
&= M^2 [s+1 + \hat{\Pi}(s) + s \delta_Z + \delta_{M^2}] \notag \\
&=: M^2 [s+1 + \hat{\Pi}^{ren}(s)], \label{eq:gluon_vertex_strict_one_loop}\\
\Gamma_{gh}^{(2)}(k_E^2) &:= - [\Delta_{gh} (k_E^2)]^{-1}  \notag \\
&= M^2 [s + \hat{\Pi}_{gh}(s) + s \delta_C] \notag \\
&=: M^2 [s + \hat{\Pi}_{gh}^{ren}(s) ], \\
\Gamma_{\psi}^{(2)}(k_E) &:= \mathcal{S}(k_E)^{-1} \notag \\
&= i \slashed{k}_E (\Gamma_{v}^{(2)}(k_E^2)  + \delta_\psi) + (\Gamma_{s}^{(2)}(k_E^2) + m_q \delta_{m_q}) \notag \\
&= i \slashed{k}_E \Gamma_{v}^{ren.}(k_E^2)  + \Gamma_{s}^{ren.}(k_E^2),
\end{align}
where $k_E$ is the Euclidean momentum, 
\begin{align}
s := \frac{k_E^2}{M^2},
\end{align}
and $\delta_Z:= Z_A - 1$, $\delta_{M^2} := Z_A Z_{M^2} -1 $, $\delta_C:= Z_C -1$, $\delta_\psi := Z_\psi -1 $, $\delta_{m_q} := Z_\psi Z_{m_q} -1$ are the counterterms.

The bare vacuum polarizations computed by the dimensional regularization read \cite{TW11,PTW14}, for gluons,
\begin{align}
\hat{\Pi}(s) &=\hat{\Pi}_{YM}(s) + \hat{\Pi}_{q}(s) \\ 
\hat{\Pi}_{YM}(s) &= \frac{g^2 C_2 (G)}{192 \pi^2} s \Biggl\{ \left( \frac{9}{s} - 26 \right) \left[ \varepsilon^{-1} + \ln (\frac{4 \pi}{M^2 e^\gamma})  \right] \notag \\
&- \frac{121}{3} + \frac{63}{s} + h(s) \Biggr\} \notag \\
\hat{\Pi}_{q}(s) &= - \frac{g^2 C(r)}{6 \pi^2} s \Biggl\{ - \frac{1}{2} \left[ \varepsilon^{-1} + \ln (\frac{4 \pi}{m_q^2 e^\gamma}) \right] \notag \\
&- \frac{5}{6} + h_q \left( \frac{\xi}{s} \right)  \Biggr\},
\end{align}
for ghosts,
\begin{align}
\hat{\Pi}_{gh}(s) &= \frac{g^2 C_2 (G)}{64 \pi^2} s \biggl[ -3 \left[ \varepsilon^{-1} + \ln (\frac{4 \pi}{M^2 e^\gamma}) \right] \notag \\
& -5 + f(s) \biggr], 
\end{align}
where $\varepsilon := 2 - D/2$, $\gamma$ is the Euler-Mascheroni constant, $C(r) = N_F/2$,
\begin{align}
\xi := \frac{m_q^2}{M^2},
\end{align}
and,
\begin{align}
h(s) &:= - \frac{1}{s^2} + \left( 1- \frac{s^2}{2} \right) \ln s \notag \\
&+ \left( 1+ \frac{1}{s}\right)^3 (s^2 - 10s + 1) \ln (s+1) \notag \\
&+ \frac{1}{2} \left( 1+ \frac{4}{s} \right)^{3/2} (s^2 - 20 s + 12) \ln \left( \frac{\sqrt{4+s} - \sqrt{s}}{\sqrt{4+s} + \sqrt{s}} \right), \notag \\
h_q(\tilde{t}) &:= 2\tilde{t} + (1-2 \tilde{t}) \sqrt{4 \tilde{t} + 1} \coth^{-1} (\sqrt{4 \tilde{t} +1}), \notag \\
   f(s) &:= - \frac{1}{s} - s \ln s + \frac{(1+s)^3}{s^2} \ln (s+1),
\end{align}
with
\begin{align}
\tilde{t} := \frac{\xi}{s} = \frac{m_q^2}{k_E^2}.
\end{align}

For the quark sector~\cite{PTW14},
\begin{align}
\Gamma_{v}^{(2)}(k_E^2) &= 1 + \Sigma_v(k_E^2), \notag \\
\Gamma_{s}^{(2)}(k_E^2) &= m_q + \Sigma_s (k_E^2)  \\
\Sigma_v(k_E^2) &= \frac{ g^2 C_2(r)}{64 \pi ^2 M^2 k_E^4}\Bigl[ K^2 \left\{ 2 M^4+M^2(k_E^2-m_q^2)\right. \notag \\
& \left.-(m_q^2+k_E^2)^2\right \} Q -2 M^2 k_E^2(-2 M^2+m_q^2 +k_E^2)  \notag \\
&-2 \{ 2 M^6+3 M^4(k_E^2-m_q^2)+(m_q^2+k_E^2)^3 \} \ln \left(\frac{m_q}{M}\right)  \notag \\
&  -2 (m_q^2+k_E^2)^3 \ln\left(\frac{m_q^2+k_E^2}{m_q^2}\right)\Bigr], \\
\Sigma_s (k_E^2) &=  - \frac{3 g^2 C_2(r) m_q}{8 \pi ^2}\Bigl[ -\frac{2}{
   \epsilon}+\ln \left(\frac{M e^{\gamma/2}}{\sqrt{4 \pi} }\right)-\frac{2}{3} \notag \\
& -\frac{K^2}{4 k_E^2} Q +\frac{1}{2 k_E^2} (M^2-m_q^2+k_E^2) \ln
   \left(\frac{m_q}{M}\right)\Bigr], \label{eq:quark-one-loop}
\end{align}
where 
\begin{align}
K^2 &:= \sqrt{M^4 + 2M^2 (k_E^2 - m_q^2) + (m_q^2 + k_E^2)^2}, \notag \\
Q &:= \ln \left( \frac{(K^2 - k_E^2)^2 - (m_q^2 - M^2)^2}{(K^2 + k_E^2)^2 - (m_q^2 - M^2)^2} \right), \label{eq:ksquared_and_q}
\end{align}
and $C_2(r) = \frac{N^2 - 1}{2N}$ for $G = SU(N)$ and fundamental quarks.

Henceforth, we adopt the ``infrared safe renormalization scheme'' by Tissier and Wschebor~\cite{TW11}, which respects the nonrenormalization theorem $ Z_A Z_C Z_{M^2} = 1$~\cite{Non-mass-ren}, in the one-loop level,
\begin{align}
 \begin{cases}
 Z_A Z_C Z_{M^2} = 1 \\
 \Gamma_{{\mathscr A}}^{(2)} (k_E = \mu) = \mu^2 + M^2\\
 \Gamma_{gh}^{(2)}(k_E = \mu) = \mu^2 \\
 \Gamma_{v}^{ren.}(\mu^2) = 1 \\
 \Gamma_{s}^{ren.}(\mu^2) = m_q
 \end{cases}
 \Leftrightarrow \ 
 \begin{cases}
 \delta_C + \delta_{M^2} = 0 \\
 \hat{\Pi}^{ren}(s = \nu) = 0 \\
 \hat{\Pi}_{gh}^{ren} (s = \nu) = 0, \\
  \Gamma_{v}^{ren.}(\mu^2) = 1 \\
 \Gamma_{s}^{ren.}(\mu^2) = m_q
 \end{cases}
\label{eq:TWrenomalization}
\end{align}
combined with the Taylor scheme \cite{Taylor71} $Z_g Z_A^{1/2} Z_C = 1$ for the coupling, where
\begin{align}
\nu := \frac{\mu^2}{M^2}.
\end{align}
In this renormalization scheme, the running coupling of some RG flows turns out to be always finite, which implies that the perturbation theory will be valid to some extent.

By imposing the above renormalization condition, we have the renormalized two-point vertex functions,
\begin{align}
\hat{\Pi}^{TW}_{ren.}(s) &=\hat{\Pi}_{YM,ren.}^{TW}(s) + \hat{\Pi}_{q,ren.}^{TW}(s), \label{eq:vacuum_pol_TW}\\
\hat{\Pi}^{TW}_{YM,ren.}(s) &= \frac{g^2 C_2 (G)}{192 \pi^2} s \biggl[ \frac{48}{s} + h(s) + \frac{3 f(\nu)}{s} - (s \rightarrow \nu) \biggr], \label{eq:vacuum_pol_TW_YM} \\
\hat{\Pi}^{TW}_{q,ren.}(s) &=  - \frac{g^2 C(r)}{6 \pi^2} s \left[ h_q \left( \frac{\xi}{s} \right) - h_q \left( \frac{\xi}{\nu} \right)  \right], \label{eq:vacuum_pol_TW_quark} \\
\hat{\Pi}_{gh,ren.}^{TW}(s) &= \frac{g^2 C_2 (G)}{64 \pi^2} s \biggl[ 
 f(s) - f(\nu) \biggr], \label{TWghostvacuumpolarization} \\
\Gamma_{v}^{TW, ren.}(k_E^2) &= 1 + \Sigma_v(k_E^2) -  \Sigma_v(\mu^2),  \\
\Gamma_{s}^{TW, ren.}(k_E^2) &= m_q + \Sigma_s(k_E^2) -  \Sigma_s(\mu^2) .
\end{align}
Note that the gluon propagator exhibits the decoupling feature and satisfies the condition (ii) of the previous section, 
\begin{align}
{\mathscr D}_T (k_E^2 = 0) = \frac{1}{M^2 [1 + \hat{\Pi}^{TW}_{ren.} (0) ]}> 0. \label{eq:pos_IR_prop_cond}
\end{align}
Indeed, 
\begin{align}
\hat{\Pi}^{TW}_{q,ren.}(s=0) &= 0, \notag \\
\hat{\Pi}^{TW}_{YM,ren.}(s=0) &= \frac{g^2 C_2 (G)}{192 \pi^2} \Biggl[ 3 f(\nu) - \frac{15}{2} \Biggr]>0,
\end{align}
where we have used $h_q(\tilde{t} \rightarrow \infty) = O(1)$, $h(s) = - \frac{111}{2s} + O(\ln s)$, $f(0) = 5/2$, and the fact that $f(s)$ is a monotonically increasing function.

Finally, note that the one-loop expression of gluon propagator will have disagreement at momenta far from the renormalization scale $\mu$. Indeed, in the UV limit, the strict one-loop expression has the wrong asymptotic form:
 \begin{align}
{\mathscr D}_T(k^2) \simeq - \left[ g^2 \gamma_0 k^2 \ln |k^2| + O(k^2)\right]^{-1}, \label{gluon_asymptotic_UV_one_loop}
\end{align}
while the RG analysis yields
\begin{align}
{\mathscr D}_T(k^2) \simeq - \frac{Z_{UV}}{k^2 (\ln |k^2|)^\gamma},
\end{align}
where we have analytically continued the gluon propagator from Euclidean momentum $k^2 = - k_E^2$ to complex $k^2$, $\gamma_0$ and $\gamma = \gamma_0/\beta_0$ are given in (\ref{eq:gamma_0_and_beta_0}), and $Z_{UV} > 0$.

However, for $\gamma = \gamma_0/\beta_0 >0$, the phase of the gluon propagator on UV momenta is qualitatively correct despite the wrong exponent of the logarithm. Furthermore, both of the propagators have negative spectral functions $\rho(\sigma^2) = \frac{1}{\pi} \operatorname{Im} {\mathscr D}_T(\sigma^2 + i \epsilon) < 0$ in the UV limit. Therefore, we expect the one-loop expression will provide a good approximation of the phase of the gluon propagator for $\gamma > 0$ based on the robustness of the winding number. In Sec.~IV, we indeed confirm that the RG improved and strict one-loop gluon propagators yield the same $N_W(C)$ for qualitatively the same parameter region.

\subsection{Renormalization group functions}

Here we present the renormalization group functions for $(g, M^2,m_q,Z_A, Z_C,Z_\psi)$~\cite{TW11,PTW14}. For later convenience, we put
\begin{align}
\lambda := \frac{C_2(G) g^2}{16 \pi^2}.
\end{align}
The beta functions $\beta_\alpha$ and anomalous dimensions $\gamma_\Phi$ are defined for the masses and coupling $\alpha = g, M^2, m_q, \lambda$ and for the fields $\Phi = {\mathscr A}, {\mathscr C}, \psi_i$ renormalized at a scale $\mu$ as
\begin{align}
\beta_\alpha := \mu \frac{d}{d\mu} \alpha,~~~ \gamma_\Phi := \mu \frac{d}{d\mu} \ln Z_\Phi,
\end{align}
where the bare masses and the bare coupling are fixed in taking the derivative.

From the nonrenormalization theorems, $Z_A Z_C Z_{M^2} = 1$ and $Z_g \sqrt{Z_A}Z_C = 1$, $\beta_\lambda$, or equivalently $\beta_g$, and $\beta_{M^2}$ are expressed by $\gamma_A$ and $\gamma_C$:
\begin{align}
\beta_\lambda = \lambda (\gamma_A + 2 \gamma_C), ~~ \beta_{M^2} = M^2 (\gamma_A + \gamma_C).
\end{align}
The other RG functions, $\gamma_C, \gamma_A, \gamma_\psi$ and $\beta_{m_q}$ are computed as follows within the one-loop approximation. By differentiating the counterterms, $\gamma_\Phi =  \mu \frac{d}{d\mu} \delta_\Phi$, we have for ghosts, \cite{TW11}
\begin{align}
\gamma_C &=  -  
 \frac{\lambda}{2 \nu^2}[2 \nu^2+2 \nu-\nu^3 \ln \nu 
\notag\\
& +(\nu-2) (\nu+1)^2 \ln (\nu+1)],
\end{align}
for gluons,\cite{TW11, PTW14}
\begin{align}
\gamma_A =& \gamma_A^{YM} + \gamma_A^{quark} \\
\gamma_A^{YM} =& - 
\frac{\lambda}{6 \nu^3}
\Biggr[ 
 \left( 17 \nu^2-74 \nu+12\right) \nu
- \nu^5 \ln \nu 
\notag\\ 
&+(\nu-2)^2 (\nu+1)^2 (2 \nu-3) \ln (\nu+1) 
\notag\\ 
&+ \nu^{3/2} \sqrt{\nu+4} \left(\nu^3-9 \nu^2+20 \nu-36\right) 
\notag\\ 
& \times
\ln \left(\frac{\sqrt{\nu+4}-\sqrt{\nu}}{\sqrt{\nu+4}+\sqrt{\nu}}\right)
 \Biggr], \notag \\
\gamma_A^{quark} =& \frac{16 \lambda  C(r)} {3 C_2(G)} \left(\frac{6 t^2 \ln \left(\frac{\sqrt{4 t+1}+1}{\sqrt{4 t+1}-1}\right)}{\sqrt{4 t+1}}-3 t+\frac{1}{2}\right),
\end{align}
where 
\begin{align}
t := \frac{m_q^2}{\mu^2},    
\end{align}
and, for quarks, \cite{PTW14}
\begin{align}
\gamma_\psi =& \frac{g^2 C_2(r)}{32 \pi ^2 \mu ^4 M^2} \Biggl\{ \frac{Q}{K^2} \Bigl[ \mu ^8+\mu ^6 \left(m_q^2+M^2\right) \notag \\
& +2 \left(M^2-m_q^2\right)^3 \left(m_q^2+2 M^2\right) \notag \\ 
& +\mu ^4 \left(-2 M^2 m_q^2-3 m_q^4+3
   M^4\right) \notag \\
& +\mu ^2 \left(M^2-m_q^2\right)  \left(6 M^2 m_q^2+5 m_q^4+7 M^4\right) \Bigr] \notag \\
& +2 \mu ^2 M^2 \left(\mu ^2-2 m_q^2+4 M^2\right) \notag \\
& +2 \left(\mu ^6-3 \mu ^2 m_q^4-2
   m_q^6\right) \ln \left(\frac{\mu ^2}{m_q^2}+1\right) \notag \\
& +2 \Bigl[ \mu ^6-3 \mu ^2 \left(m_q^4+M^4\right) \notag \\
&-2 \left(-3 M^4 m_q^2+m_q^6+2 M^6\right)\Bigr] \ln \left(\frac{m_q}{M}\right) \Biggr\}.
\end{align}
Since $Z_{m_q} = 1 + \delta_{m_q} - \delta_\psi$ in the one-loop level, we obtain the beta function for the quark mass $m_q$ \cite{PTW14},
\begin{align}
\beta_{m_q} = m_q \gamma_\psi +
\frac{3 g^2 C_2(r) m_q }{16 \pi ^2} \Bigl[ -2 +\frac{2
   \left(M^2-m_q^2\right) }{\mu ^2}\ln \left(\frac{m_q}{M}\right) \notag \\
 -\frac{\left(\mu ^2 \left(m_q^2+M^2\right)+\left(M^2-m_q^2\right)^2\right) Q }{\mu^2 K^2} \Bigr],
\end{align}
where $Q$ and $K^2$ are given in (\ref{eq:ksquared_and_q}) with $k_E^2 = \mu^2$.

 \begin{figure}[tb]
  \begin{center}
   \includegraphics[width=0.8\linewidth]{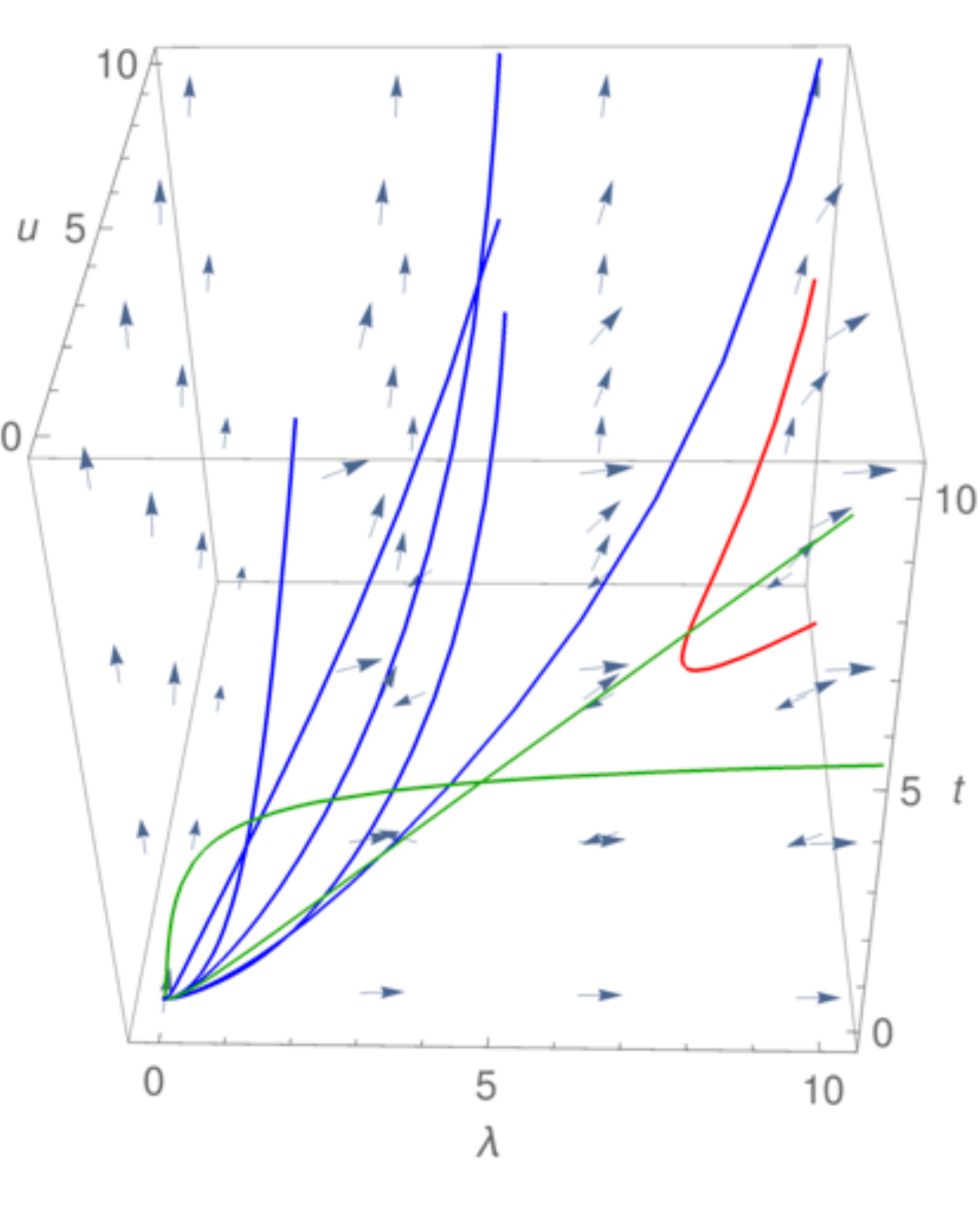}
  \end{center}
   \caption{RG flows in the parameter space $(\lambda= \frac{C_2(G) g^2}{16 \pi^2},u = M^2/\mu^2, t = m_q^2/\mu^2)$ for $N_F = 3$. The arrows indicate infrared directions $\mu \rightarrow 0$. The blue trajectories are infrared safe. The green ones end at an infrared Landau pole. The red one is not ultraviolet asymptotic free.}
    \label{fig:flowdiagramnf3}
\end{figure}

The flow diagram of $(\lambda= \frac{C_2(G) g^2}{16 \pi^2},u = M^2/\mu^2, t = m_q^2/\mu^2)$ for $N_F = 3$ is depicted in Fig.~\ref{fig:flowdiagramnf3}. In Fig.~{\ref{fig:flowdiagramnf3}}, the infrared safe trajectories are shown as the blue curves, while the green curves terminate at an infrared Landau pole. The red one is not ultraviolet asymptotic free in the sense that the running coupling $\lambda(\mu)$ does not vanish in the UV limit $\mu \rightarrow \infty$ on the RG flow. As the pure Yang-Mills case, the diagram has the infrared safe trajectories, on which the running gauge coupling $\lambda$ is always finite in the all scales. One can confirm that the qualitative feature is the same for $N_F \leq 16$.

\subsection{Asymptotic behaviors of infrared safe trajectories}

From the flow diagram Fig. \ref{fig:flowdiagramnf3}, the infrared safe trajectories possess the following features: (i) In the UV limit $\mu \rightarrow \infty$, the parameters $(\lambda,u,t)$ tend to $(\lambda \rightarrow 0, u \rightarrow 0, t \rightarrow 0)$, (ii) In the IR limit $\mu \rightarrow 0$, they tend to $(\lambda \rightarrow 0, u \rightarrow \infty, t \rightarrow \infty)$. In this subsection, we study their asymptotic behaviors within the one-loop level.

Beforehand, there is a caveat on this discussion. In the RG analysis of Oehme and Zimmermann~\cite{OZ80}, the parameter of the theory is only the gauge coupling $g$, which guarantees that the higher-loop effects are suppressed in the ultraviolet region precisely. Then, the one-loop RG gives a strong argument on asymptotic behaviors and enables us to establish the UV negativity of the gluon spectral function for $N_F < 10$. However, the massive Yang-Mills model has the mass parameter, which can potentially invalidate the perturbation theory in $\lambda$ even though $\lambda \rightarrow 0$ asymptotically. Here, we assume that the perturbation theory ``works well''. In particular, we assume that the spectral function is dominated by the one-loop contribution in both UV and IR limits.

In the pure Yang-Mills case, a similar discussion on the RG functions and Euclidean propagators can be found in e.g.,~\cite{RSTW17}. In what follows, we study asymptotic behaviors of the RG functions, Euclidean propagators, and spectral functions in the massive Yang-Mills model described in the previous subsections.
\subsubsection{UV limit}

We consider the UV limit and confirm that the asymptotic behaviors of the massive Yang-Mills model is consistent with those of the Faddeev-Popov Lagrangian. Taking the limit $u,t \rightarrow 0$, we have
\begin{align}
\gamma_A &\rightarrow  \left( - \frac{13}{3} + \frac{8 C(r)}{3C_2 (G)} \right) \lambda + O(u,t),\\
\gamma_C &\rightarrow - \frac{3}{2} \lambda +O(u \ln u),
\end{align}
which reproduce the standard one-loop beta function:
\begin{align}
\beta_\lambda &= \lambda (\gamma_A + 2 \gamma_C) \rightarrow  \beta_{0,\lambda} \lambda^2, \\
\beta_{0,\lambda} &:= - \frac{22}{3} + \frac{8 C(r)}{3C_2 (G)} < 0,
\end{align}
recovering the UV asymptotic behavior of the gauge coupling $\lambda \sim - \beta_{0,\lambda} / \ln(\mu/\Lambda)$ with some constant scale $\Lambda$. On the other hand, the beta functions for the masses read, for gluons,
\begin{align}
\beta_{M^2} &= M^2 (\gamma_A + \gamma_C) \rightarrow  \beta_{0,M^2} \lambda M^2, \notag \\
\beta_{0,M^2} &:= - \frac{35}{6} + \frac{8 C(r)}{3C_2 (G)},
\end{align}
and for quarks,
\begin{align}
\beta_{m_q} &\rightarrow  - \frac{6 C_2 (r)}{C_2(G)} \lambda m_q + O(u\ln u, t \ln t).
\end{align}
Notice that at $G = SU(3)$, the gluon mass is suppressed logarithmically for $\beta_{0,M^2} <0$, or $N_F < \frac{35}{8} C_2(G) = 105/8 \approx 13.1$ and enhanced logarithmically for $14 \leq N_F \leq 16$, while the correction to the quark mass always suppresses the quark mass in the logarithmic way.

Note that $u = M^2/ \mu^2 \rightarrow 0$ and $t = m_q^2/\mu^2 \rightarrow 0$ exponentially faster than $\lambda \rightarrow 0$. This justifies \textit{a posteriori} taking the limit $u,t \rightarrow 0$ in the first step.

Next, let us consider the propagators on the Euclidean momenta. From the nonrenormalization theorems, $Z_A = Z_\lambda Z_{M^2}^{-2}$ and $Z_C = Z_{M^2}Z_\lambda^{-1}$, which yield together with the renormalization conditions~(\ref{eq:TWrenomalization}),~\cite{TW11},
\begin{align}
\mathscr{D}_T (k_E^2,\mu_0^2) &= \frac{\lambda_0}{M_0^4} \frac{M^4(k_E^2)}{\lambda(k_E^2)} \frac{1}{k_E^2 + M^2(k_E^2)}, \notag \\
\Delta (k_E^2,\mu_0^2) &= -\frac{M_0^2}{\lambda_0} \frac{\lambda(k_E^2)}{M^2(k_E^2)} \frac{1}{k_E^2}, \label{eq:Euc_gluon_ghost}
\end{align}
where $\mu_0$ is the renormalization scale, and $M_0$ and $\lambda_0$ are the mass and coupling at $\mu_0$. In the UV limit $k_E^2 \rightarrow \infty$,
\begin{align}
\mathscr{D}_T (k_E^2,\mu_0^2) &\sim [\lambda(k_E^2)]^{\frac{2 \beta_{0,M^2}}{\beta_{0,\lambda}} -1} \frac{1}{k_E^2} \sim \frac{1}{k_E^2 (\ln k_E^2)^{\frac{\gamma_{0,A}}{\beta_{0,\lambda}}}} , \notag \\
\Delta (k_E^2,\mu_0^2) &\sim  - [\lambda(k_E^2)]^{- \frac{\beta_{0,M^2}}{\beta_{0,\lambda}} +1} \frac{1}{k_E^2} \sim -\frac{1}{k_E^2 (\ln k_E^2)^{\frac{\gamma_{0,C}}{\beta_{0,\lambda}}}}, \label{eq:UV_asymptotic}
\end{align}
in accordance with the asymptotic form~\cite{OZ80}, where we have defined $\gamma_{0,A} :=  - \frac{13}{3} + \frac{8 C(r)}{3C_2 (G)}  = 2 \beta_{0,M^2} - \beta_{0,\lambda}$ and $\gamma_{0,C} := - \frac{3}{2}  =  \beta_{0,\lambda} - \beta_{0,M^2}$. These lead to the superconvergence relations for the gluon spectral function when $\gamma_{0,A}< 0$, and also for the ghost spectral function. For quarks, one can find $\gamma_\psi = O(\lambda^2,u,t)$, which gives no logarithmic correction to the quark propagator.

One can see the UV negativity for the gluon spectral function and UV positivity for the ghost spectral function from (\ref{eq:UV_asymptotic}) combined with the analyticity or RG improvement with respect to $|k^2|$ in the complex $k^2$ plane~\cite{OZ80}.

Thus, it turns out that the massive Yang-Mills model can describe the correct UV behavior by RG improvement from the above observations, although we have regarded the massive Yang-Mills model as a low-energy effective model.

\subsubsection{IR limit}

Next, we consider the IR limit of the infrared safe trajectories: $\lambda \rightarrow 0, u \rightarrow \infty, t \rightarrow \infty$. Taking the limit $u,t \rightarrow \infty$, we have
\begin{align}
\gamma_A &\rightarrow \frac{1}{3} \lambda + O (u^{-1}, t^{-1}), \notag \\
\gamma_C &\rightarrow 0 + O(u^{-1} \ln u), \label{eq:asymptotic_gluon_ghost_anomalous_dimension}
\end{align}
as those of the pure Yang-Mills case shown in~\cite{RSTW17}. This indicates that the infrared gluon and ghost sector is insensitive to the quark flavor effect except the case of ``massless quark'' $m_q = 0$. As shown in~\cite{RSTW17}, we find  $\beta_\lambda/\lambda = \beta_{M^2}/M^2 = \lambda/3$, which leads as $\mu \rightarrow 0$ to
\begin{align}
M^2 \sim \lambda \sim \frac{1}{\ln \mu^{-1}}. \label{eq:infrared_coupling}
\end{align}
The quark mass $m_q$ runs according to
\begin{align}
\beta_{m_q} &\rightarrow  O(u^{-1} \ln u, u^{-1}  \ln t, t^{-1}), \label{eq:infrared_quark_mass_beta}
\end{align}
which indicates absence of logarithmic correction to the quark mass in the infrared.
Since $u = M^2/ \mu^2 \rightarrow \infty$ and $t = m_q^2/\mu^2 \rightarrow \infty$ exponentially increase more rapidly than $\lambda \sim \frac{1}{\ln \mu} \rightarrow 0$, the above evaluation is consistent.

From (\ref{eq:infrared_coupling}) and (\ref{eq:Euc_gluon_ghost}), the infrared safe trajectories describe the decoupling solution, i.e., the massive gluon and the massless free ghost in the IR limit as those in the pure massive Yang-Mills model of~\cite{RSTW17}.

Finally, we examine the spectral functions. In the gluon vacuum polarization, the quark loop affects the gluon spectral function in the region of $(2 m_q)^2 < k^2 <\infty$ as $h_q(\tilde{t})$ has the branch cut only on $-1/4<\tilde{t}<0$. Therefore, the IR spectra of the gluon and ghost are the same as those in the pure Yang-Mills case within the one-loop level because the quark mass will be finite in the infrared  as can be seen from (\ref{eq:infrared_quark_mass_beta}).

As $s \rightarrow 0$, the vacuum polarizations are
\begin{align}
\hat{\Pi}^{TW}_{YM,ren.}(s) &= \frac{\lambda}{12} \biggl[ 3 f(\nu) - \frac{15}{2} + s \ln s + O(s) \biggr], \notag \\
\hat{\Pi}_{gh,ren.}^{TW}(s) &= \frac{\lambda}{4} s \biggl[ \left( \frac{5}{2} - f(\nu) \right) - s \ln s + O(s^2)  \biggr]. 
\end{align}

From the assumption that the higher loop terms are suppressed as $\mu \rightarrow 0$, i.e., the running $u$ and $t$ do not invalidate the perturbation theory in $\lambda$, the following approximation holds in the IR limit $|k^2| \rightarrow 0$,
\begin{align}
\mathscr{D}_T (k^2,\mu_0^2) &\simeq Z_A (|k^2|,\mu_0^2) \mathscr{D}_T^{1-loop} (k^2,|k^2|) , \notag \\
\Delta (k^2,\mu_0^2) &\simeq Z_C (|k^2|,\mu_0^2) \Delta^{1-loop} (k^2,|k^2|) \label{eq:RG_infrared_complex_ansatz}
\end{align}
where $Z_A (|k^2|,\mu_0^2)$ and $Z_C (|k^2|,\mu_0^2)$ are the renormalization factors and $Z_A (\mu^2,\mu_0^2) = \frac{\lambda_0}{M_0^4} \frac{M^4(\mu^2)}{\lambda(\mu^2)}$, $Z_C (\mu^2,\mu_0^2) = \frac{M_0^2}{\lambda_0} \frac{\lambda(\mu^2)}{M^2(\mu^2)}$ from the nonrenormalization theorems. As $\mu^2 \rightarrow 0$,
\begin{align}
Z_A  (\mu^2,\mu_0^2) \sim M^2(\mu^2), ~~ Z_C (\mu^2,\mu_0^2) \sim \mathrm{const.}
\end{align}
Therefore, the spectral functions originating from the branch cut on timelike momenta have the IR asymptotic forms $\sigma^2 \rightarrow +0$: for gluons,
\begin{align}
\rho(\sigma^2) &= \frac{1}{\pi} \operatorname{Im} \mathscr{D}_T (\sigma^2 + i \epsilon,\mu_0^2) \notag \\
&\simeq Z_A (\sigma^2,\mu_0^2) \frac{1}{\pi} \operatorname{Im} \mathscr{D}_T^{1-loop} (\sigma^2 + i \epsilon,\sigma^2) \notag \\
&\sim \operatorname{Im} \left. \frac{1}{ \left[ 1 + s + \hat{\Pi}^{TW}_{ren.}(s) \right]} \right|_{s = - \frac{\sigma^2}{M^2} - i \epsilon} \notag \\
& \sim - \operatorname{Im} \hat{\Pi}^{TW}_{ren.}(- \frac{\sigma^2}{M^2} - i \epsilon) \notag \\
&\sim - \frac{\lambda(\sigma^2)}{M^2(\sigma^2)} \sigma^2  \sim - \sigma^2 < 0, \label{eq:infrared_negativity_gluon}
\end{align}
and for ghosts,
\begin{align}
\rho_{gh}(\sigma^2) &= \frac{1}{\pi} \operatorname{Im} \Delta (\sigma^2 + i \epsilon,\mu_0^2) \notag \\
&\simeq Z_C (\sigma^2,\mu_0^2) \frac{1}{\pi} \operatorname{Im} \Delta^{1-loop} (\sigma^2 + i \epsilon,\sigma^2) \notag \\
&\sim - \operatorname{Im} \left. \frac{1}{ M^2(\sigma) \left[ s + \hat{\Pi}^{TW}_{gh,ren.}(s) \right]} \right|_{s = - \frac{\sigma^2}{M^2} - i \epsilon} \notag \\
&\sim \mathrm{const.} > 0.
\end{align}
These results demonstrate \textit{the IR negativity of the gluon spectral function} and \textit{the IR positivity of the ghost spectral function} for the infrared safe trajectories with $m_q > 0$. The ghost spectral function has a delta function at $\sigma^2 = 0$ with a negative coefficient associated to the negative norm massless ghost state and shows the finite positive spectrum in the limit $\sigma^2 \rightarrow + 0$. Although we assume that the running $u$ and $t$ do not invalidate the asymptotic perturbative expansion in $\lambda$, the IR negativity of the gluon spectral function will be a remarkable consequence from the infrared safety.

As an alternative evidence for the IR negativity of the gluon spectral function, we can utilize the IR behavior of the Euclidean propagator. For example, although this relation gives a trivial result in this model, it is worthwhile to note that they are related as $\lim_{k_E \rightarrow + 0} \frac{d}{d k_E} \mathscr{D}_T (k_E^2) = - \pi \lim_{\sigma \rightarrow + 0} \frac{d}{d \sigma} \rho(\sigma^2)$~\cite{CPRW18}. Let us consider the IR asymptotic behavior of the gluon propagator. The Euclidean propagator (\ref{eq:Euc_gluon_ghost}) can be rewritten as~\cite{RSTW17},
\begin{align}
\mathscr{D}_T (k_E^2)  = \frac{Z_C^{-1} (k_E^2)}{1 + k_E^2/M^2(k_E^2)},
\end{align}
from which we have
\begin{align}
k_E^2 \frac{d}{d k_E^2}\mathscr{D}_T (k_E^2)  &\simeq \frac{1}{2} Z_C^{-1} (k_E^2) [- \gamma_C - 2 u^{-1}] \notag \\
&\simeq \frac{1}{2} Z_C^{-1} (k_E^2)  u^{-1} [\frac{\lambda}{2} u^{-1} \ln u - 2] \notag \\
&\sim - k_E^2 \ln k_E^2 > 0,
\end{align}
where we have used $\gamma_C \rightarrow - \frac{\lambda}{2} u^{-1} \ln u + O(\lambda u)$ and $\lambda(\mu) \simeq - \frac{6}{\ln \mu^2}$ as $\mu \rightarrow 0$. Therefore, as $k_E \rightarrow 0$, we have
\begin{align}
\frac{d}{d k_E^2 }\mathscr{D}_T (k_E^2) \sim |\ln k_E^2| \rightarrow + \infty. 
\end{align}
This logarithmic divergence is precisely consistent with the IR spectral negativity of $\rho(\sigma^2) \sim - \sigma^2 < 0$. Indeed, if $\rho(\sigma^2) \sim - \sigma^2 < 0$, then, from the (generalized) spectral representation, as $k_E \rightarrow +0$,
\begin{align}
\frac{d}{d k_E^2 }\mathscr{D}_T (k_E^2) &\simeq - \int_0^\infty d \sigma^2 ~ \frac{\rho(\sigma^2)}{(\sigma^2 + k_E^2)^2} \notag \\
&\sim |\ln k_E^2|,
\end{align}
which shows the coherency between the approximation of (\ref{eq:RG_infrared_complex_ansatz}) in the complex $k^2$ momentum and the IR asymptotic behavior of the Euclidean propagator of (\ref{eq:Euc_gluon_ghost}). Both of the two arguments imply the IR negativity of the gluon spectral function. As negativity of a spectral function in a weak sense leads to complex poles \cite{HK2018}, the IR and UV negativity of the gluon spectral function supports the existence of complex poles in the gluon propagator.

For the quark propagator, the one-loop expression gives no nontrivial IR spectrum from (\ref{eq:quark-one-loop}). This is rather clear from the facts that the Feynman diagram indicates $\operatorname{Im} \Sigma $ has nonvanishing value only for $k^2 > m_q^2$ and that $m_q(\mu)$ is constant in the IR limit.

Incidentally, let us comment on the nontrivial infrared fixed point to give the Gribov-type scaling solution~\cite{RSTW17}. In the massive Yang-Mills model of the pure Yang-Mills theory, a nontrivial fixed point in $(\lambda,u)$ plane appears, where the gluon and ghost exhibit the Gribov-type behavior. One can find a similar fixed point in the massive Yang-Mills model with ``massless quarks'' $m_q = 0$. For instance, the fixed point lies at $(\lambda,u,t) = (4.0,1.6,0)$ in $N_F = 3$.
However, this scaling-type fixed point has an infrared unstable direction. Therefore, with quarks $m_q > 0$, the unstable scaling-type fixed point disappears due to the running of the quark mass. 

\section{Numerical results on analytic structures}

In this section, we present results on the number of complex poles of the gluon and quark propagators in the massive Yang-Mills model with quarks at the one-loop approximation and its RG improvements. Without quarks, we found the gluon propagator has $N_P = - N_W(C) = 2$ for all parameters $(g^2,M^2)$ due to the negativity of the spectral function \cite{HK2018}. Surprisingly, it turns out that the model with many light quarks has $N_P = 0,2,4$ regions, depending on the number of quarks $N_F$ and the parameters $(g^2,M^2,m_q^2)$. In particular, for $N_F~(4 \lesssim N_F \leq 9)$ light quarks, the $N_P = 4$ region, where the gluon propagator has two pairs of complex poles, covers a typical value of coupling $g$, e.g., $g \sim 4$ in the setting to be described shortly, and the models with $N_F~(N_F \geq 10)$ very light quarks have the region with no complex poles around the typical value of coupling $g$.

From here on, we set $G = SU(3)$ and the renormalization scale $\mu_0 = 1 ~\mathrm{GeV}$. With the RG improvements, the best fit parameters reported in \cite{PTW14} are $g = 4.5,
M = 0.42$ GeV, and the up and down quark masses $m_{u} = m_{d} = 0.13$ GeV for the case of $N_F = 2$ while $g = 5.3, M = 0.56$ GeV, and $m_{u} = m_{d} = 0.13$ GeV for $N_F = 2 + 1 + 1$ (with the assumption on the strange and charm quark masses of $m_{s} = 2 m_u$ and $m_c = 20 m_u$). Notice that the ``quark mass'' $m_q$ of this model should not be confused with the current quark mass. The ``quark mass'' parameter $m_q$ will be chosen to reproduce the propagators. Rather, $m_q$ will be of the same order of the constituent quark mass, since we renormalized this model at $\mu_0 = 1$ GeV.
In particular, the massless quarks will differ from the ``massless quarks'' $m_q = 0$ due to the spontaneous breakdown of the chiral symmetry.

First, we investigate $N_W(C)$ of the strict one-loop gluon propagator, mainly at fixed typical values of the parameters $g = 4$ and $M^2 = 0.2~ \mathrm{GeV}^2$ to see a qualitative overview of the analytic structures of this model. Next, we consider the RG improvements of these results and evaluate the gluon spectral function at the best-fit parameters for $N_F = 2$. Comparing the strict one-loop and RG-improved results could support the robustness of $N_W(C)$. Furthermore, we discuss the complex poles of the quark propagator and comment on the ghost propagator.

\subsection{Gluon propagator: Strict-one-loop analysis}

 \begin{figure}[tb]
  \begin{center}
   \includegraphics[width=0.9\linewidth]{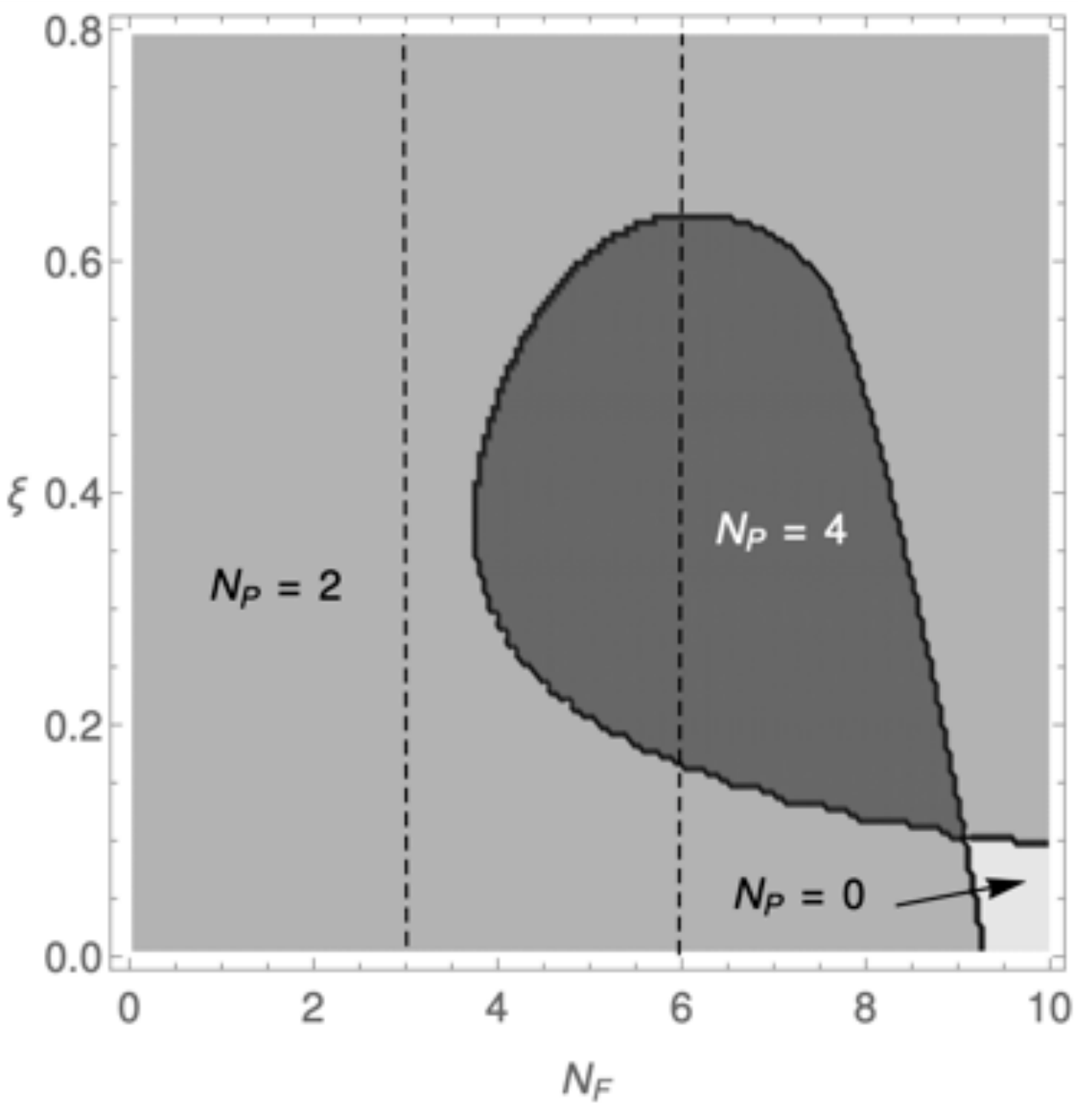}
  \end{center}
   \caption{Contour plot of $N_W(C)$ for the gluon propagator on the $(N_F,\xi)$ plane at $g = 4$ and $M^2 = 0.2~ \mathrm{GeV}^2$, which gives the number of complex poles through the relation $N_P = - N_W(C)$. In the $N_P = 0,2,4$ regions, the gluon propagator has zero complex pole, one pair, and two pairs of complex conjugate poles, respectively. The vertical dashed lines at $N_F = 3$ and $N_F = 6$ correspond to the top and bottom figures of Fig.~{\ref{fig:flavor_pole_location_B}}, respectively.}
    \label{fig:flavor_quasiposineg_windingnumber}
\end{figure}

Based on the strict one-loop gluon propagator (\ref{eq:gluon_vertex_strict_one_loop}) combined with (\ref{eq:vacuum_pol_TW}), (\ref{eq:vacuum_pol_TW_YM}), and (\ref{eq:vacuum_pol_TW_quark}) of Sec.~III, we compute the winding number $N_W(C)$ for $N_F \leq 9$ according to the procedure (\ref{eq:winding-timelike}) of Sec.~II. Notice that the strict one-loop gluon propagator satisfies the conditions (i) and (ii) in Sec.~II. Since the two-point vertex function $\Gamma_{{\mathscr A}}^{(2)} = [{\mathscr D}_T ]^{-1}$ is finite, the gluon propagator has no zeros, $N_Z = 0$. Thus we can count the number of complex poles by computing $N_W(C) = - N_P$.

\subsubsection{The number of complex poles at the typical $(g,M)$}
As a first step, we investigate the $(N_F,\xi = m_q^2/M^2)$ dependence of $N_P$ at the fixed typical values of the parameters to obtain an overview.

Figure \ref{fig:flavor_quasiposineg_windingnumber} displays the contour plot of $N_W(C)$ on the $(N_F,\xi = \frac{m_q^2}{M^2})$ plane at the fixed $g = 4$ and $M^2 = 0.2~ \mathrm{GeV}^2$. This figure is restricted to $0 \leq N_F < 10$, since the one-loop approximation will be not reliable for $N_F \geq 10$. Indeed, the naive one-loop UV asymptotic form (\ref{gluon_asymptotic_UV_one_loop}) for $\gamma_0 > 0$, i.e., for $N_F \geq 10$, shows the existence of a Euclidean pole together with the fact that $\mathscr{D}_T (0) > 0$ and $\mathscr{D}_T(k^2)$ has no zeros. 

This result illustrates that the gluon propagator has four complex poles in the region colored with dark gray ($4 \lesssim N_F \leq 9$, $0.2 \lesssim \xi \lesssim 0.6$). It predicts that the transition occurs from the $N_P = 2$ region to $N_P = 4$ region by adding light quarks since $N_W(C)$ is a topological invariant. In the $N_P = 4$ region, the gluon propagator has two pairs of complex conjugate poles if we exclude the possibility of Euclidean poles on the negative real axis.

To see details of the $N_P = 4$ region and its boundary, let us look into positions of complex poles.

\subsubsection{Location of complex poles}

Here we take a further look into positions of complex poles and the transition of $N_W(C)$. We will report the ratio $w/v$ for a position of a complex pole $k^2 = v + iw$ and trajectories of complex poles for varying $\xi$.

First, we focus on the ratio $w/v$ at the complex pole $k^2 = v + i w$, on which we set $w \geq 0$ without loss of generality from the Schwarz reflection principle, in order to see whether or not the gluon presents a particlelike resonance. Since the gluon propagator has at most two pairs of complex conjugate poles, it is sufficient to find $\max w/v$ and $\min w/v$.

\begin{figure}[t]
 \begin{minipage}{\hsize}
  \begin{center}
   \includegraphics[width=0.8\linewidth]{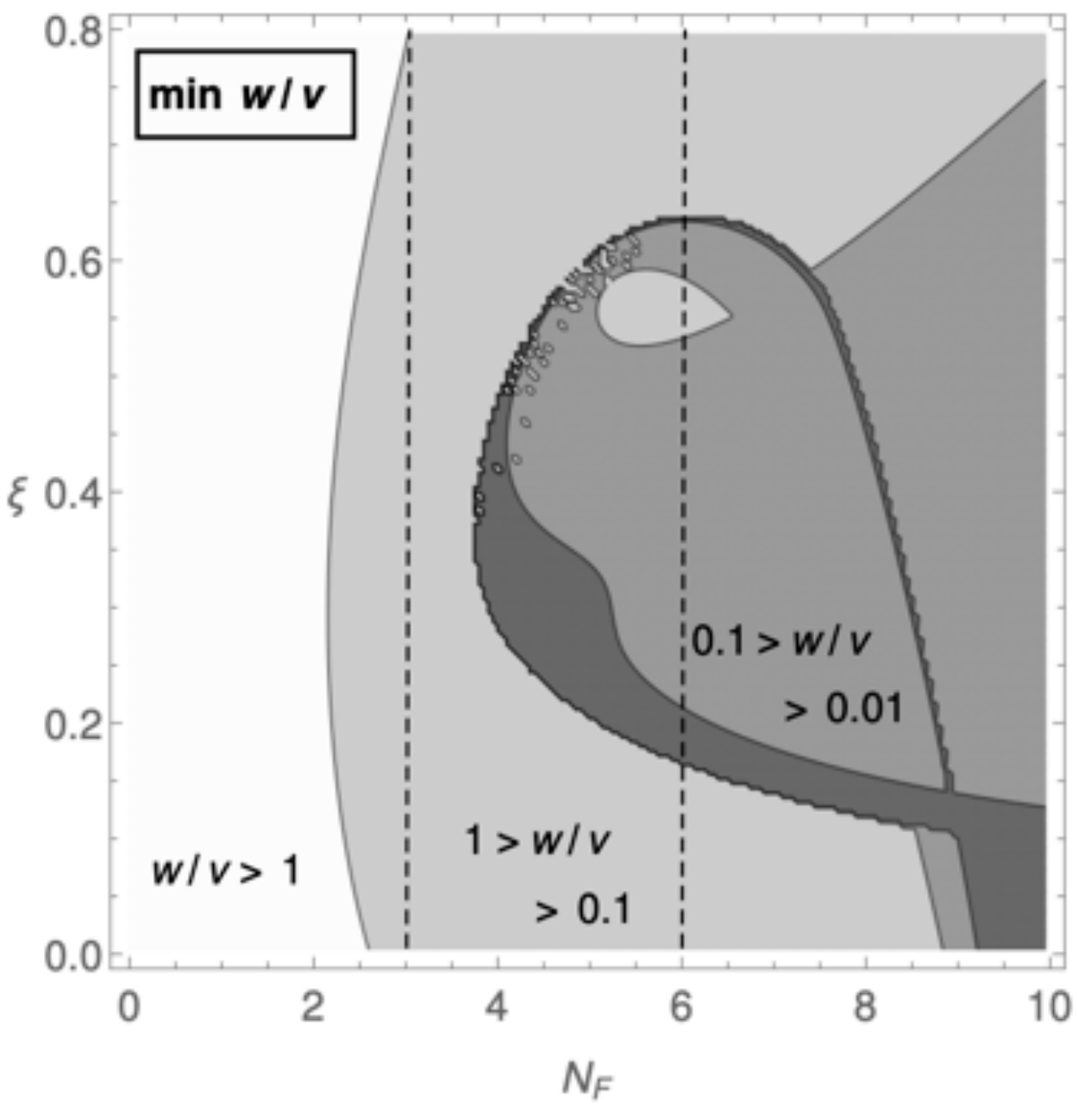}
  \end{center}

 \end{minipage}
 \begin{minipage}{\hsize}
  \begin{center}
   \includegraphics[width=0.8\linewidth]{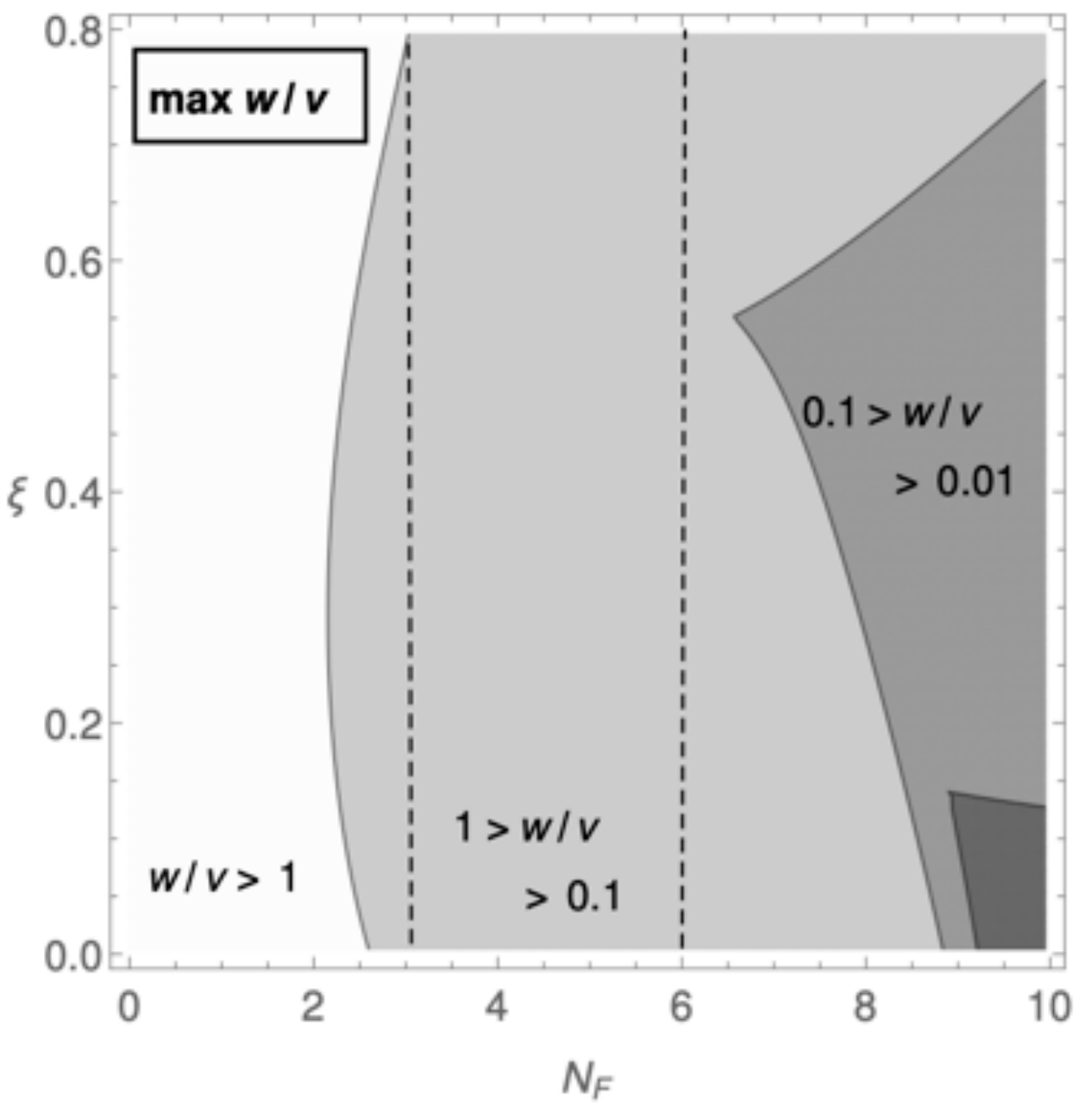}
  \end{center}
 \end{minipage}
 \caption{Contour plots of $\operatorname{min} w/v$ (top) and $\operatorname{max} w/v$ (bottom) for a complex pole at $k^2 = v + i w,~ w \geq 0$ of the gluon propagator on the $(N_F,\xi)$ plane. 
 The regions of $ w/v>1$, $1> w/v>0.1$, $0.1 > w/v>0.01$, and $0.01 > w/v$ are represented by different levels of gray. The vertical dashed lines at $N_F = 3$ and $N_F = 6$ correspond to the top and bottom figures of Fig.~{\ref{fig:flavor_pole_location_B}}, respectively. The region of $\min w/v < 0.01$ appears around the transition between $N_W(C) = -2$ and $N_W(C) = -4$. As $N_F$ increases, the ratio $w/v$ tends to be small.}
    \label{fig:flavor_pole_location_A}
\end{figure}

The results are demonstrated in Fig.~\ref{fig:flavor_pole_location_A}. In general, the ratio $w/v$ of a complex pole decreases as $N_F$ increases. This implies that the gluon exhibits more particlelike behavior for larger $N_F$. Note that the ratio $\min w/v$ rapidly decreases near the boundary between $N_W(C) = -2$ and $N_W(C) = -4$. 

\begin{figure}[tb]
 \begin{minipage}{\hsize}
  \begin{center}
   \includegraphics[width=0.9\linewidth]{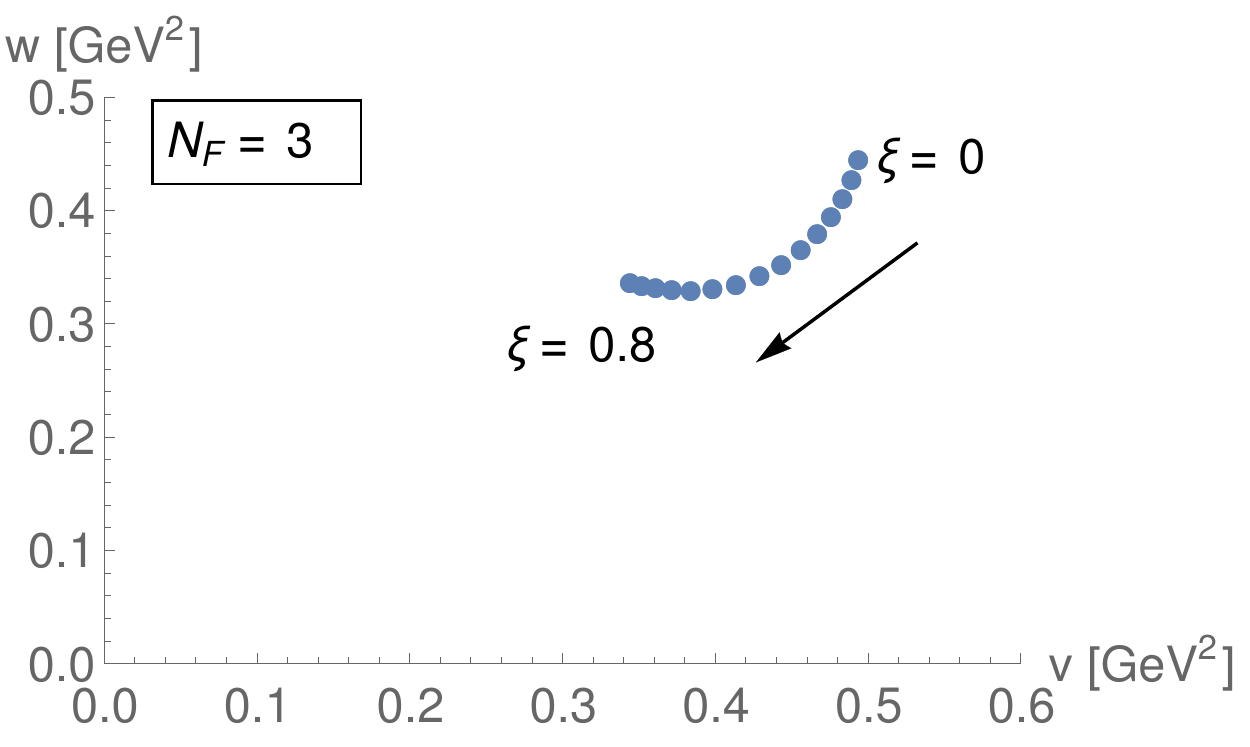}
  \end{center}

 \end{minipage}
 \begin{minipage}{\hsize}
  \begin{center}
   \includegraphics[width=0.9\linewidth]{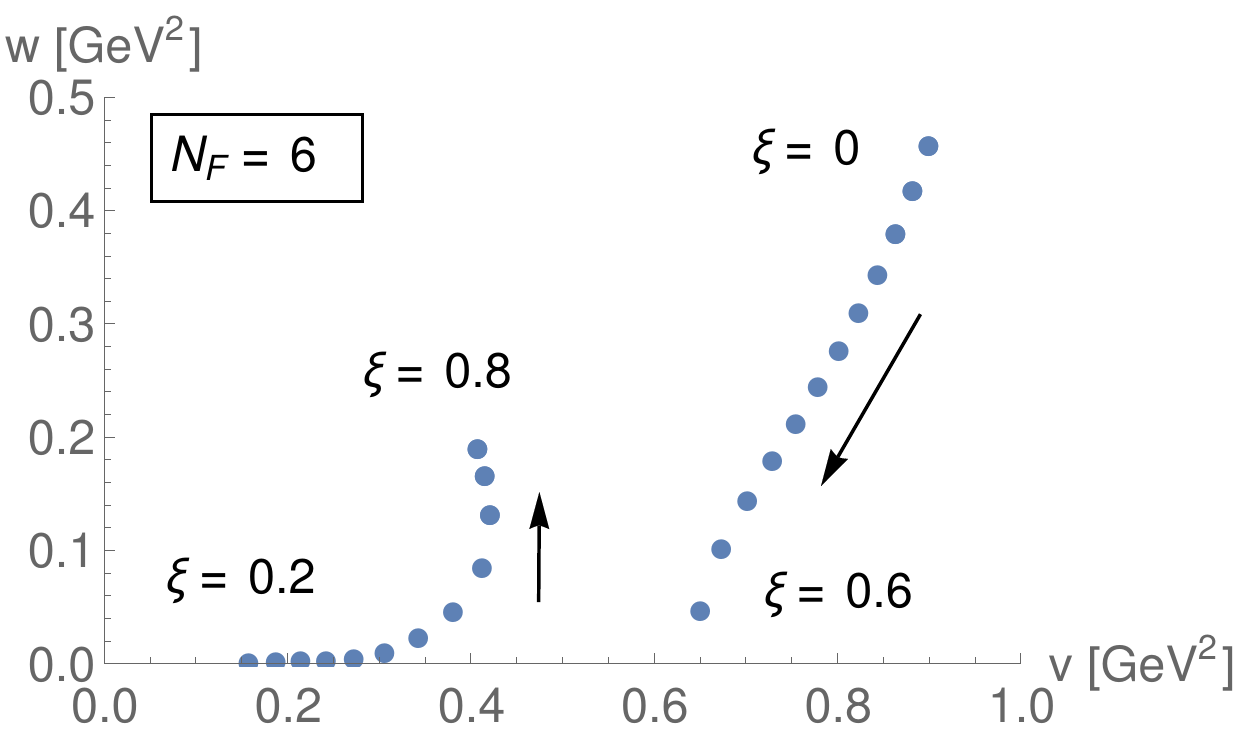}
  \end{center}
 \end{minipage}
 \caption{Positions of poles of the gluon propagator at $N_F = 3$ (top) and $N_F = 6$ (bottom) for varying $0<\xi =\frac{m_q^2}{M^2} < 0.8$. As $\xi$ increases, the poles move in the direction shown by the arrows. At $N_F = 3$ (top), the pole moves continuously from the position of $\xi = 0$. In contrast, at $N_F = 6$ (bottom), the pole from the position of $\xi = 0$ goes behind the branch cut at $\xi \approx 0.6$. Moreover, the new pole ($v \approx 0.16$ GeV$^2$) appears from the branch cut at $\xi \approx 0.2$. }
    \label{fig:flavor_pole_location_B}
\end{figure}

Second, we examine trajectories of complex poles for $N_F = 3$ and $N_F = 6$. Figure \ref{fig:flavor_pole_location_B} plots the pole location with varying $\xi$. In the case of $N_F = 3$, no transition changing $N_P$ occurs; the pole moves gradually and lowers its real part $v$ as increasing $\xi$. On the other hand, the trajectory of complex poles is completely different at $N_F = 6$ where the transition occurs. The pole moves continuously from the position of $\xi = 0$, but is absorbed into the branch cut at $\xi \approx 0.6$. Beforehand, the new pole arises at $\xi \approx 0.2$ from the branch cut.

These observations demonstrate that (i) the gluon propagator has a pole at timelike momentum, which could correspond to a physical particle, on the boundary between the $N_P =2$ and $N_P = 4$ regions and that (ii) the pole bifurcates and becomes a new pair of complex conjugate poles in the $N_P = 4$ region. This appearance of the new pair of complex conjugate poles from the branch cut on the real positive axis is compatible with the rapid decrease of $\operatorname{min} w/v$ on the boundary between the $N_P = 2$ and $N_P = 4$ regions in Fig.~\ref{fig:flavor_pole_location_A}.

\subsubsection{(g,M) dependence}

We have seen that the gluon propagator changes its number of complex poles when many light quarks are incorporated at the fixed typical values of the parameters of $g = 4$ and $M^2 = 0.2~ \mathrm{GeV}^2$. Therefore, we have to check whether or not the existence of the transition is insensitive to choices of the parameters $(g,M)$. 

One can verify that the $N_W(C) = -4$ region in the parameter space $(\lambda = \frac{C_2(G) g^2}{16 \pi^2}, u = M^2/\mu_0^2, \xi= \frac{m_q^2}{M^2})$ largely expands by changing $N_F$ from $N_F = 3$ to $N_F = 6$. At $N_F = 6$, the $N_W(C) = -4$ region dominates the parameter region around the typical value $g \sim 4$, $M^2 \sim 0.2~ \mathrm{GeV}^2$ with $0.2 \lesssim \frac{m_q^2}{M^2} \lesssim 0.6$. See Appendix B for details.

Thus our qualitative conclusion will be valid: this model has the transition of $N_P$, and the gluon propagator has four complex poles in the presence of $N_F ~ (4 \lesssim N_F \leq 9)$ light quarks with $m_q$ satisfying $0.2 \lesssim \frac{m_q^2}{M^2} \lesssim 0.6$.

\begin{figure}[tb]
 \begin{minipage}{\hsize}
  \begin{center}
   \includegraphics[width=0.8\linewidth]{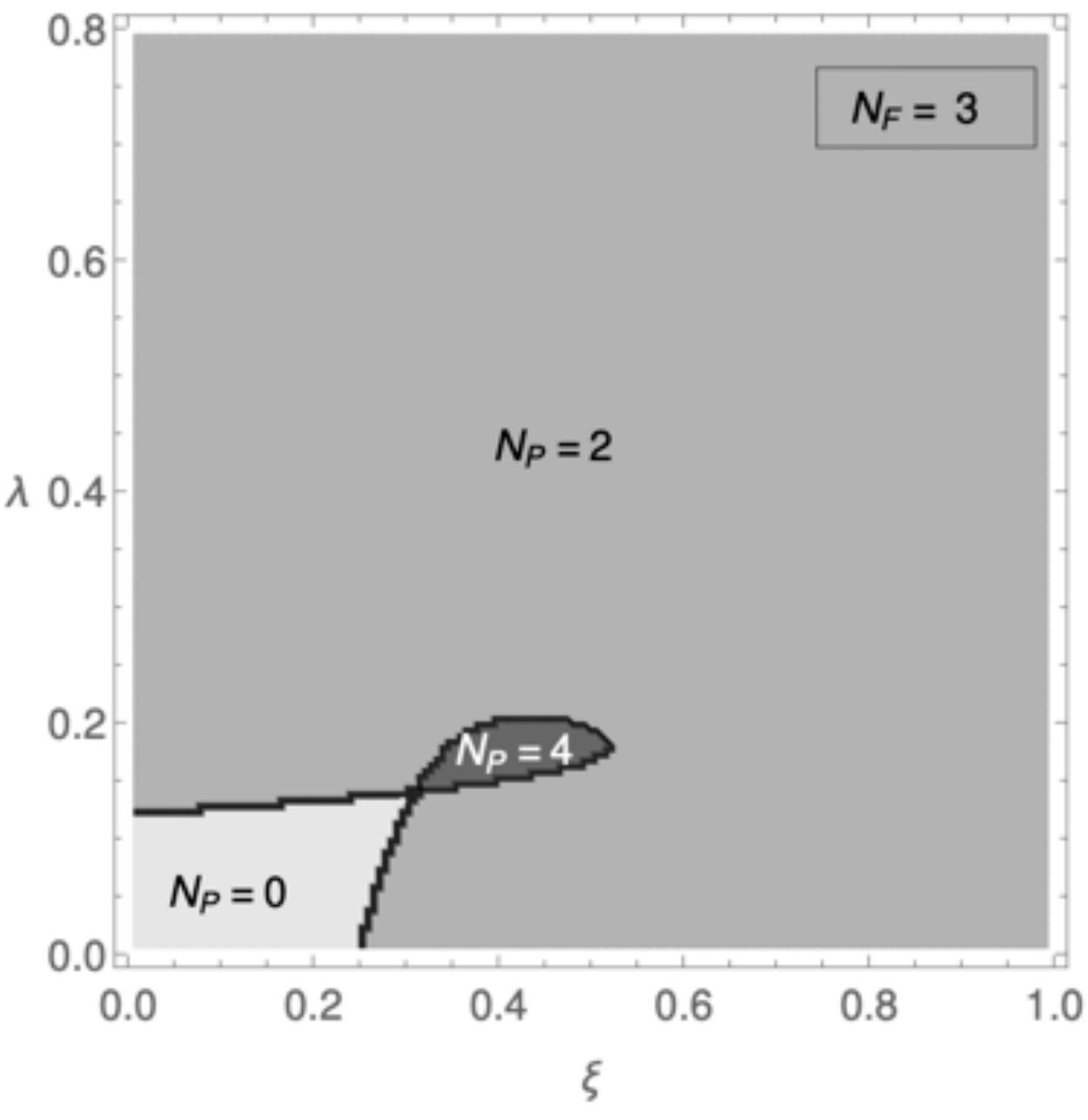}
  \end{center}

 \end{minipage}
 \begin{minipage}{\hsize}
  \begin{center}
   \includegraphics[width=0.8\linewidth]{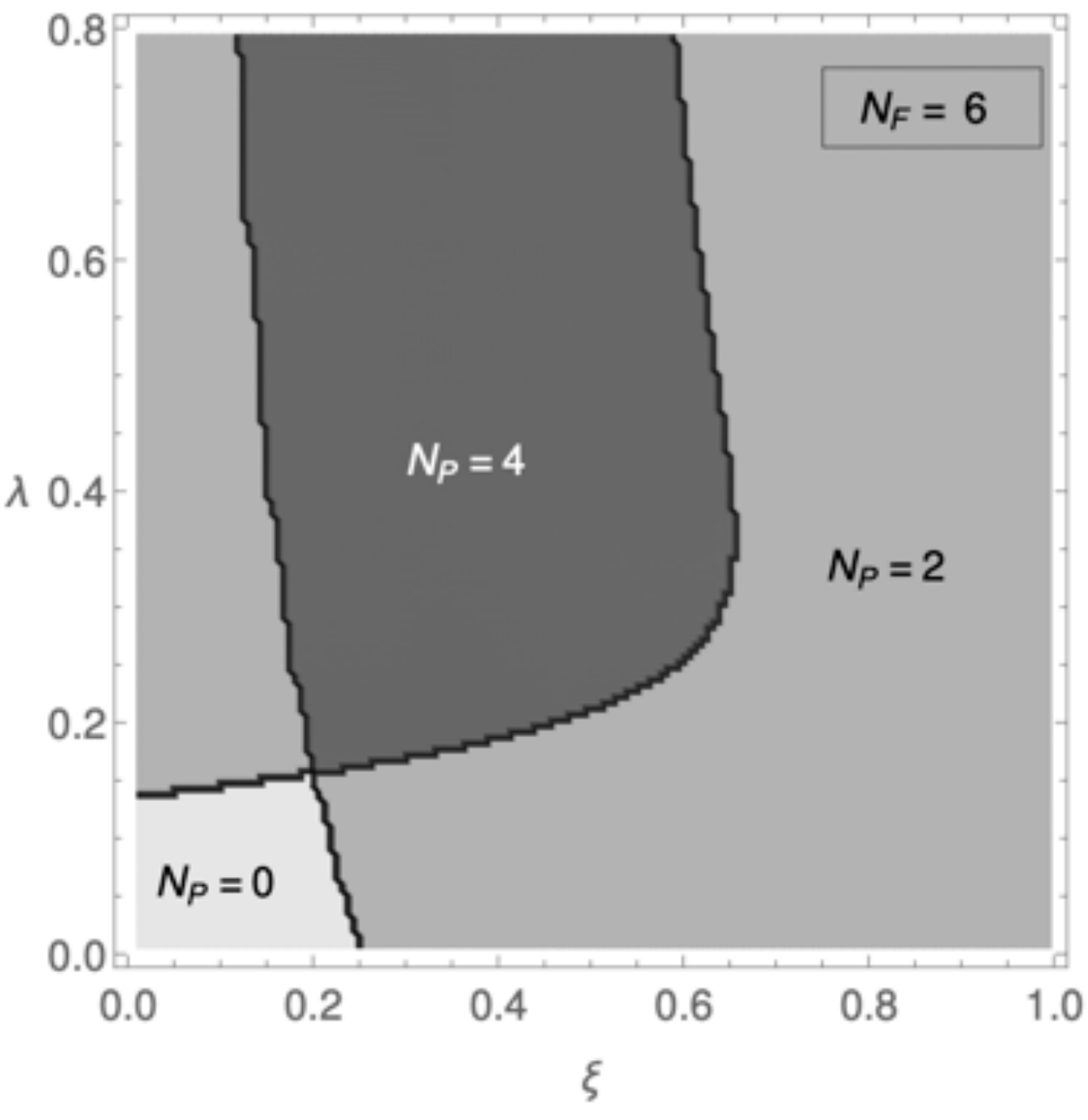}
  \end{center}
 \end{minipage}
 \caption{Two-dimensional slice of equi-$N_W(C)$ volume of the strict one-loop gluon propagator at $N_F = 3$ (top) and $N_F = 6$ (bottom) in the $(\lambda = \frac{C_2(G) g^2}{16 \pi^2}, u = M^2/\mu_0^2 = 0.2, \xi= \frac{m_q^2}{M^2})$ space.}
    \label{fig:strict-one-loop-nf36}
\end{figure}

Incidentally, to compare the results with those of the RG improved gluon propagator, we show in Fig. \ref{fig:strict-one-loop-nf36} contour plots of $N_W(C)$ at $u = M^2/\mu_0^2 = 0.2$.

\subsection{Gluon propagator: RG improved analysis}

So far, we have studied the gluon propagator in the strict one-loop level, relying on the robustness of the winding number. To make this robustness more reliable and to study the structure of the gluon propagator for $N_F \geq 10$, it is important to survey the winding number for the RG improved gluon propagator. Moreover, this model reproduces the numerical lattice results with the RG improved propagators at the ``realistic'' parameters $g = 4.5,
M = 0.42$ GeV, and $m_{q} = 0.13$ GeV at $\mu_0 = 1$ GeV and $N_F = 2$ \cite{PTW14}. In this subsection, we investigate the analytic structure of the one-loop RG improved gluon propagator for the realistic parameters and its parameter dependence for each number of quark flavors $N_F$.

 \begin{figure}[tb]
  \begin{center}
   \includegraphics[width=\linewidth]{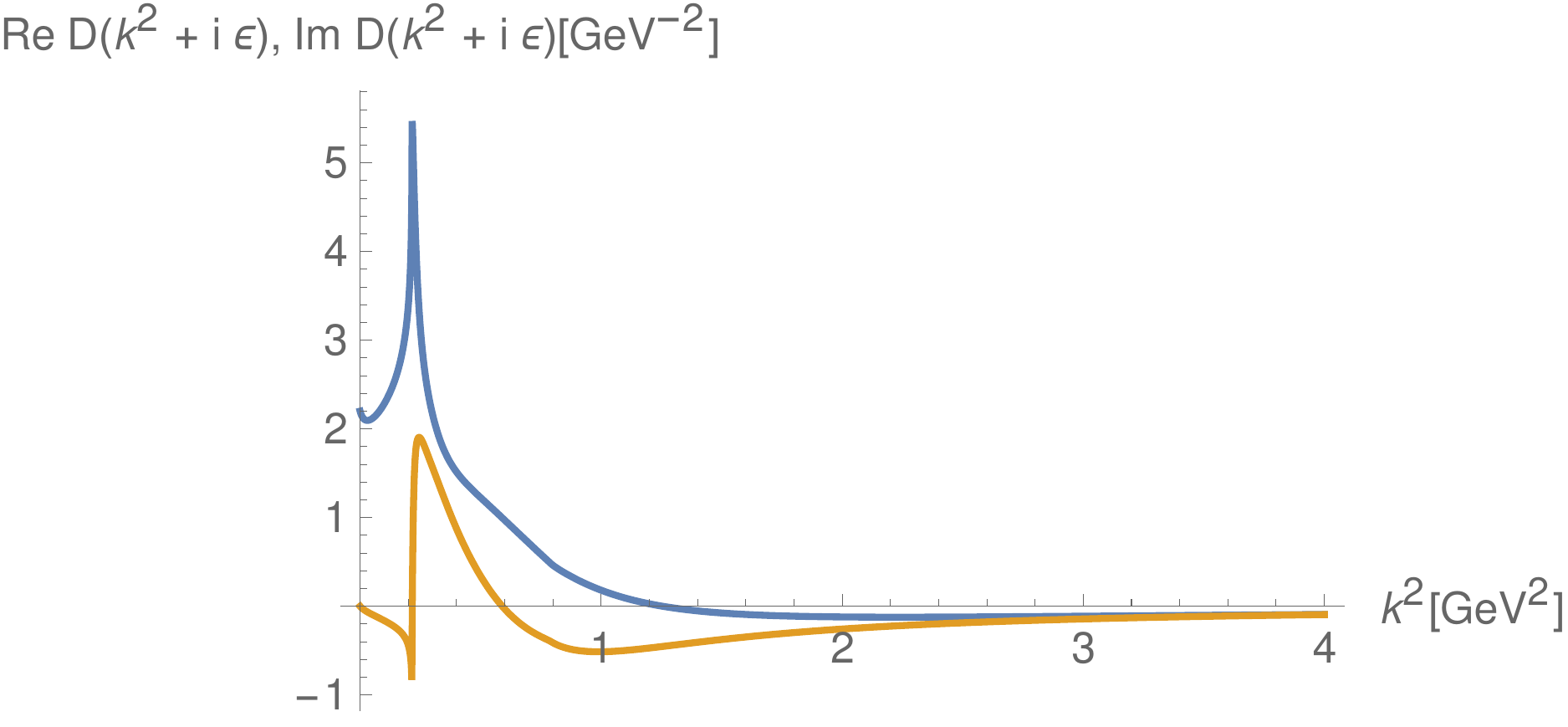}
  \end{center}
 \caption{Real part (blue) and imaginary part (orange) of the gluon propagator at the realistic parameters for $N_F = 2$ \cite{PTW14} on the positive real axis. This shows its spectral function is quasinegative, from which $N_W(C) = -2$. Notice that the spectral function exhibits the linear decrease with respect to $k^2$ in agreement with (\ref{eq:infrared_negativity_gluon}) in the IR limit.}
    \label{fig:realistic-gluon}
\end{figure}

The RG equation for the gluon propagator is
 \begin{align}
\mathscr{D}_T (k^2_E,\alpha(\mu^2),\mu^2) &= Z_A^{-1} (\mu^2,\mu_0^2) \mathscr{D}_T (k^2_E,\alpha(\mu_0^2),\mu_0^2)
\end{align}
 where $\alpha$ denotes the set of gauge coupling and masses $\alpha = (\lambda,u=M^2/\mu^2,t=m_q^2/\mu^2)$ and $Z_A (\mu^2,\mu_0^2)$ is the renormalization factor computed by the anomalous dimension. 
We then approximate the gluon propagator to avoid the large logarithms as
 \begin{align}
\mathscr{D}_T &(k^2,\alpha(\mu_0^2),\mu_0^2) \approx \notag \\
&Z_A (|k^2|,\mu_0^2) \mathscr{D}_T^{1-loop} (k^2,\alpha(|k^2|),|k^2|) \label{eq:RG_improvement_scheme}
\end{align}
Although the RG improvements only for the modulus $|k^2|$ on the complex $k^2$ plane may break the analyticity, this will give a better approximation than the strict one-loop approximation for the propagator on the timelike momenta.

First, let us see the gluon propagator for the realistic parameters.
Figure~\ref{fig:realistic-gluon} plots the real and imaginary parts of the gluon propagator on the timelike momenta. The spectral function is quasinegative, i.e., the spectral function is negative $\rho(k_0^2) < 0$ at all timelike zeros $k_0^2$ of ${\rm Re}\ D(k^2)$. Then, $N_W(C) = -2$ from the invariance of $N_W(C)$ under continuous deformation \cite{HK2018}. Since the gluon propagator has no zeros $N_Z = 0$ from the fact that the RG improvement only affects the renormalization factor and the running couplings, we deduce the gluon propagator of this model has one pair of complex conjugate poles as in the pure Yang-Mills case. Incidentally, notice that the spectral function $\rho(\sigma^2)$ decreases linearly with respect to $\sigma^2$ in the IR limit $\sigma^2 \rightarrow +0$, which is a general feature for the infrared safe trajectories with $m_q > 0$ in this model as shown in (\ref{eq:infrared_negativity_gluon}).

 \begin{figure}[tb]
 \begin{minipage}{\hsize}
  \begin{center}
   \includegraphics[width=0.8\linewidth]{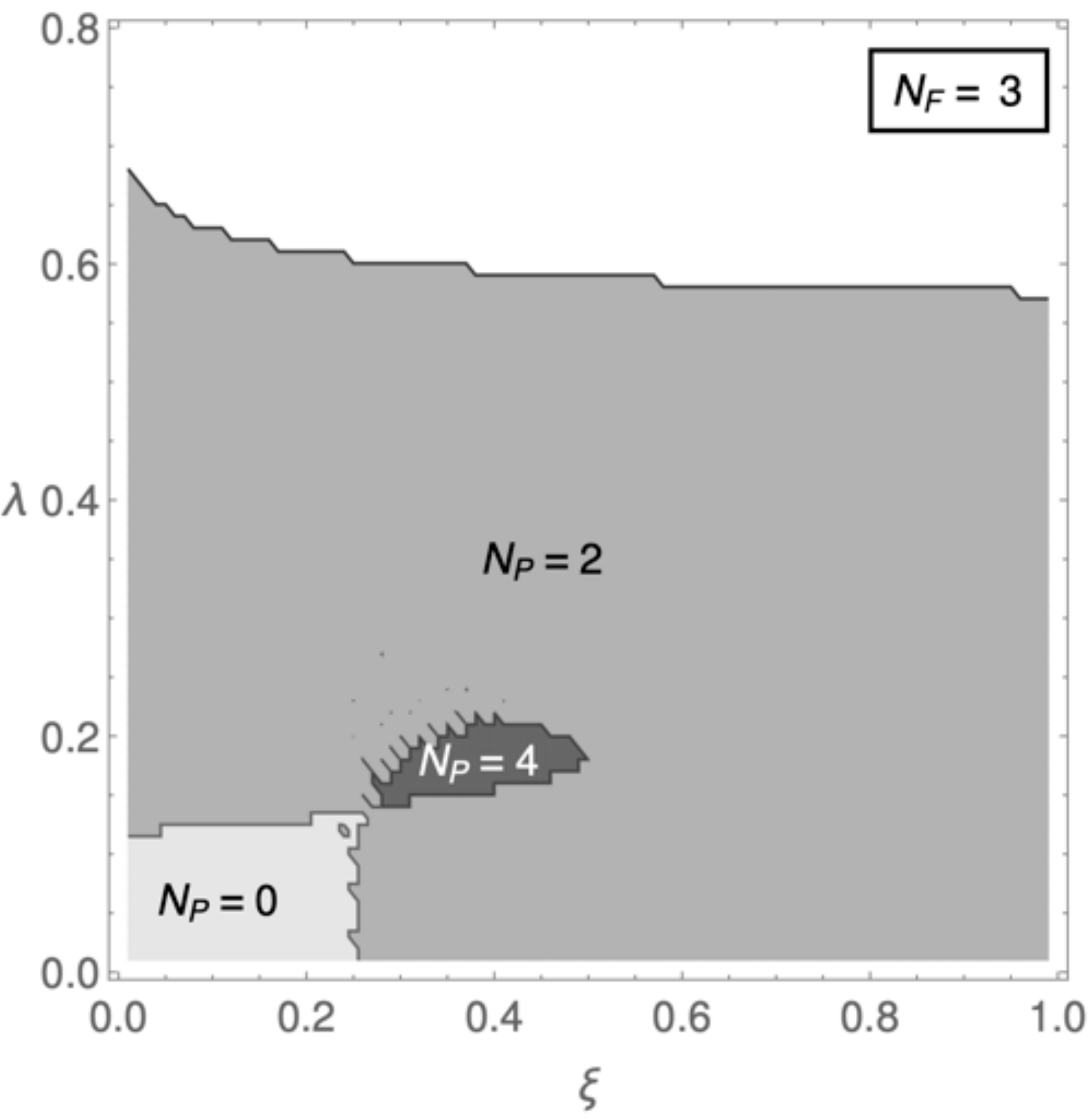}
  \end{center}

 \end{minipage}
 \begin{minipage}{\hsize}
  \begin{center}
   \includegraphics[width=0.8\linewidth]{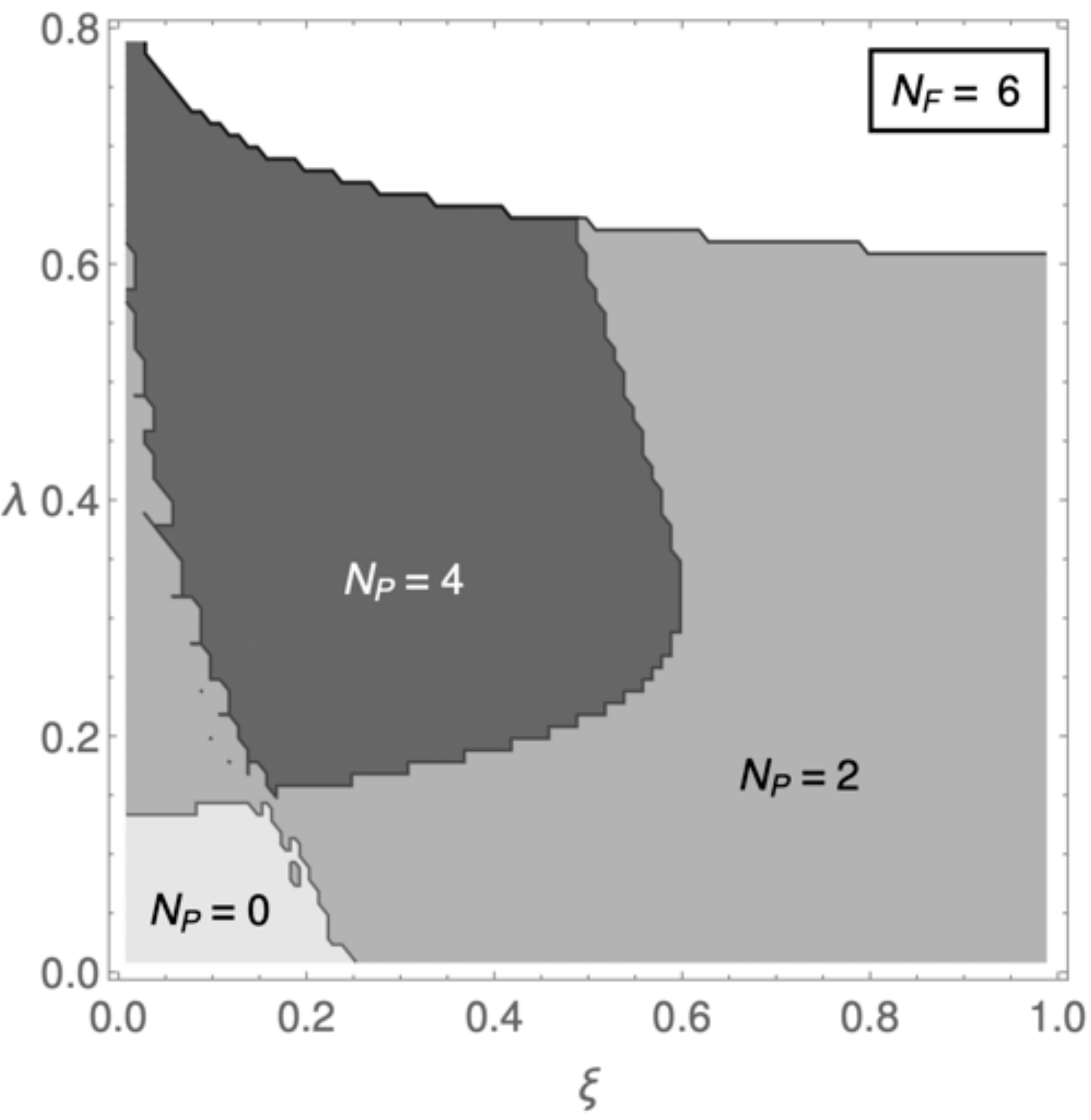}
  \end{center}
 \end{minipage}
 \caption{Two-dimensional slice of equi-$N_W(C)$ volume of the gluon propagator at $N_F = 3$ (top) and $N_F = 6$ (bottom) in the $(\lambda = \frac{C_2(G) g^2}{16 \pi^2}, u= M^2/\mu_0^2 = 0.2, \xi= \frac{m_q^2}{M^2})$ space. Besides the region with the Landau poles (white), these plots have almost the same structure as Fig.~\ref{fig:strict-one-loop-nf36}.}
    \label{fig:RG-one-loop-nf36}
\end{figure}

 \begin{figure}[tb]
 \begin{minipage}{\hsize}
  \begin{center}
   \includegraphics[width=0.8\linewidth]{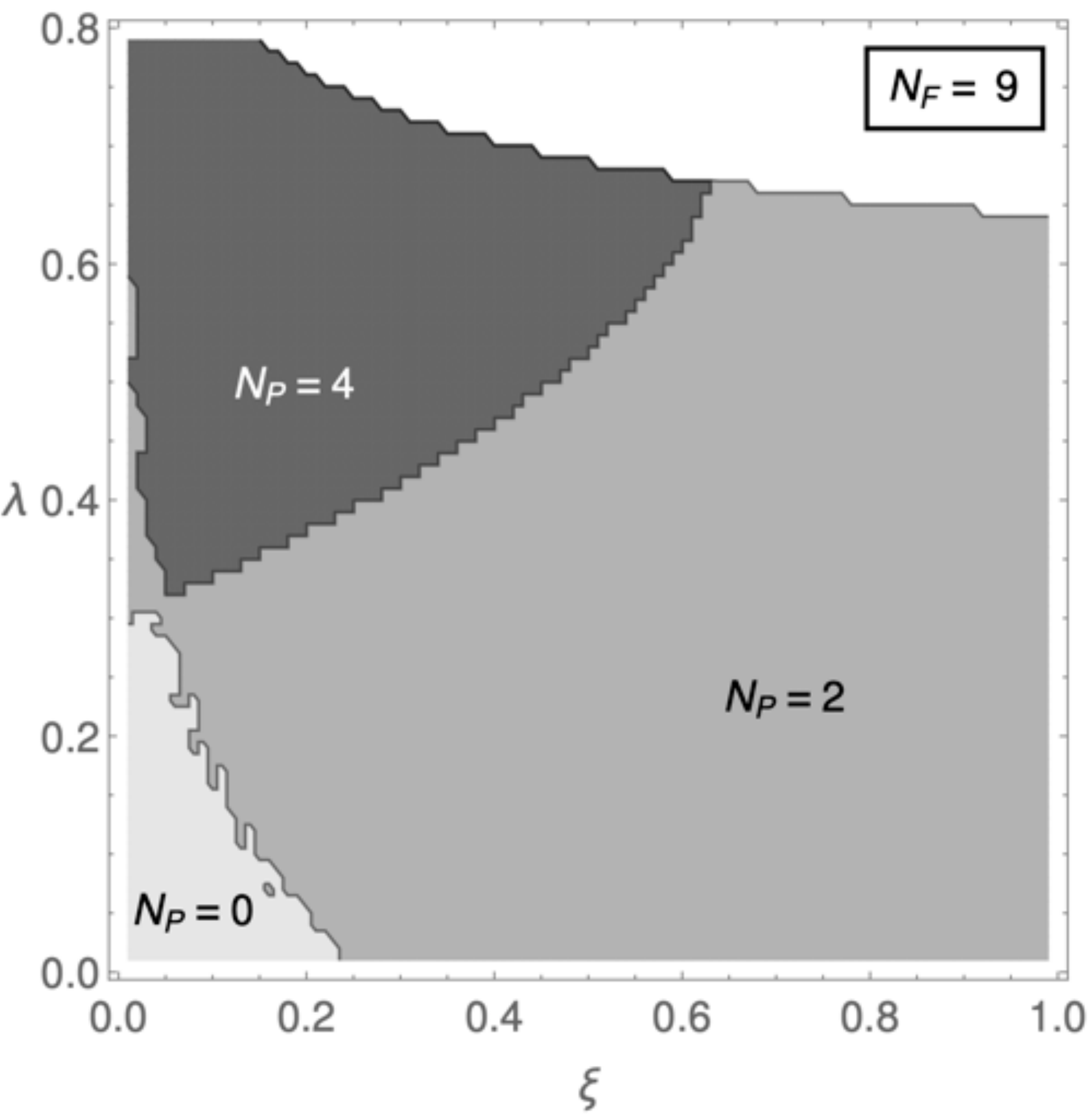}
  \end{center}

 \end{minipage}
 \begin{minipage}{\hsize}
  \begin{center}
   \includegraphics[width=0.8\linewidth]{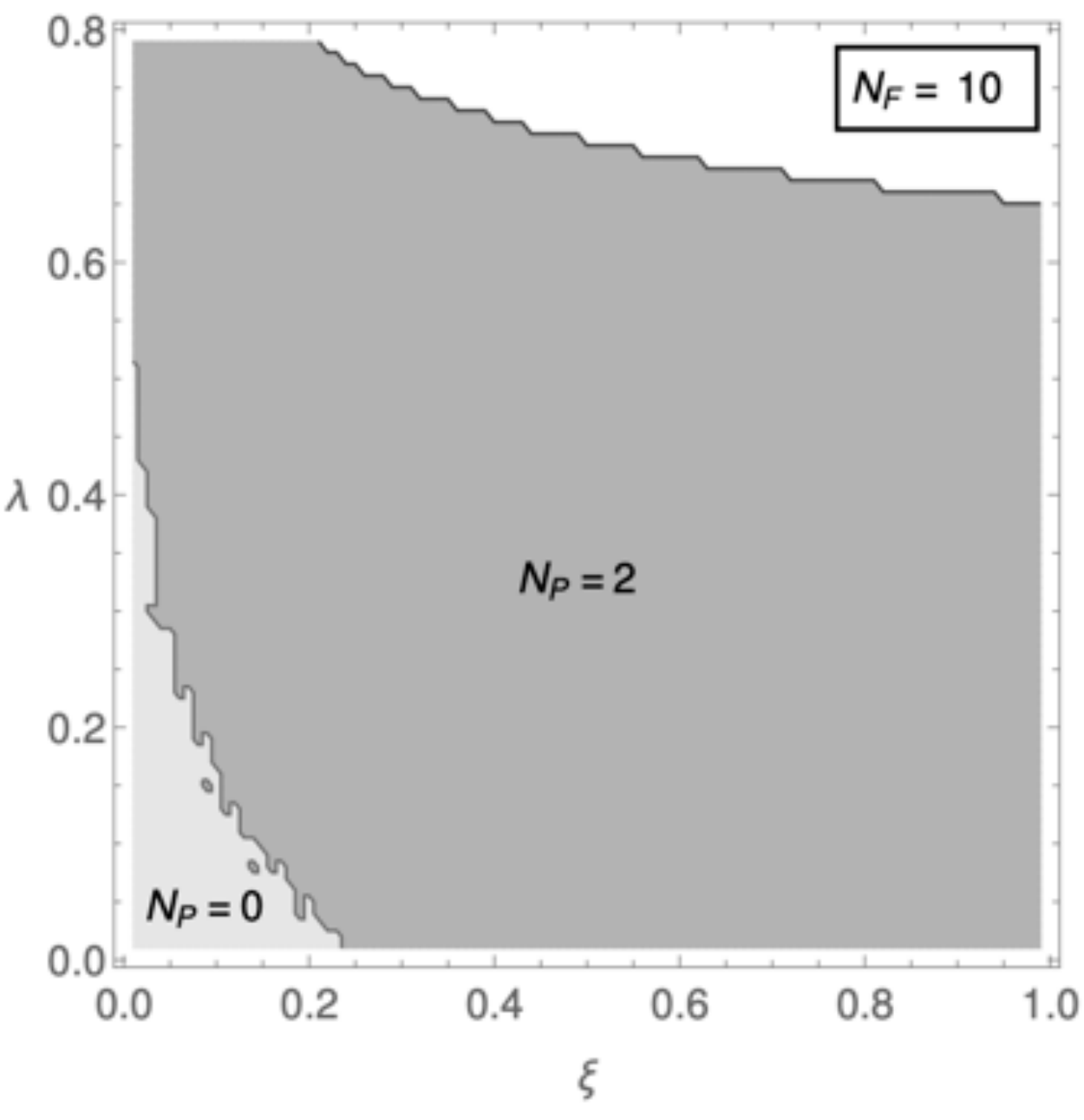}
  \end{center}
 \end{minipage}
 \caption{Two-dimensional slice of equi-$N_W(C)$ volume of the gluon propagator at $N_F = 9$ (top) and $N_F = 10$ (bottom) in the $(\lambda = \frac{C_2(G) g^2}{16 \pi^2}, u= M^2/\mu_0^2 = 0.2, \xi= \frac{m_q^2}{M^2})$ space. The white region shows RG trajectories with the Landau poles.}
    \label{fig:RG-one-loop-nf910}
\end{figure}

Next, we investigate the number of complex poles $N_P$ in the whole parameter space for each $N_F$. Since $N_Z = 0$ within this approximation as before, $N_P$ can be computed as $N_P = -N_W(C)$. Note that different points on a same renormalization group trajectory in the three dimensional parameter space, e.g., Fig.~\ref{fig:flowdiagramnf3}, provide the same analytic structure. Indeed, they are exactly connected by the scale transformation from the dimensional analysis and the anomalous dimension: under a scaling $\kappa$,
 \begin{align}
\mathscr{D}_T &(k^2,\alpha(\kappa^2 \mu_0^2),\mu_0^2) \notag \\
&= \kappa^2 Z_A^{-1} (\kappa^2 \mu_0^2,\mu_0^2) \mathscr{D}_T (\kappa^2 k^2,\alpha(\mu_0^2),\mu_0^2),
\end{align}
which shows that $\mathscr{D}_T (k^2,\alpha(\kappa^2 \mu_0^2),\mu_0^2)$ and $\mathscr{D}_T (k^2,\alpha(\mu_0^2),\mu_0^2)$ have the same number of complex poles.
Therefore, it suffices to compute in the two-dimensional slice of the renormalization group flow, see Fig.~\ref{fig:flowdiagramnf3}. From here on, we employ the two dimensional slice at $u = M^2/\mu^2 = 0.2$.

Figures \ref{fig:RG-one-loop-nf36} and \ref{fig:RG-one-loop-nf910} reveal the results for $N_F = 3,6,9,10$. Note that, since the analytic structure of the gluon propagator approaches to that of the pure massive Yang-Mills model as $m_q \rightarrow \infty$, the gluon propagator has one pair of complex conjugate poles $N_P = 2$ for large $\xi$.
At $N_F = 3$, the $N_P = 2$ region dominates the region of the parameters. At $N_F = 6$, in the presence of light quarks, the $N_P = 4$ region, on which the gluon propagator has two pairs of complex conjugate poles, occupies the larger region, especially around a typical coupling $\lambda \sim 0.3$. On the other hand, at $N_F = 9$, the $N_P = 0$ region, on which the gluon propagator has no complex poles, expands and the $N_P = 4$ region shrinks. At $N_F = 10$, notably, the $N_P = 4$ region disappears, and the $N_P = 0$ region appears in the larger region of $\lambda$ for the very light quark mass $m_q/M \ll 1$.

Note that the strict one-loop gluon propagator and the RG improved one provide almost the same result for $N_F = 3,6$ by a comparison between Fig. \ref{fig:strict-one-loop-nf36} and Fig. \ref{fig:RG-one-loop-nf36}. This supports the robustness of the winding number $N_W(C)$.

It is remarkable that the parameter dependence of the analytic structure drastically changes between $N_F = 9$ and $N_F = 10$. The extension of $N_P = 0$ region for very light quark mass can be seen from the UV positivity of the spectral function for $N_F \geq 10$~\cite{OZ80}, since the positivity in a weak sense indicates $N_W(C) = 0$ \cite{HK2018}. In a similar sense, the disappearance of $N_P = 4$ region could be related to the UV positivity of the spectral function, which might lower $N_W(C)$.

\subsection{Quark propagator}

As pointed out in~\cite{PTW14}, the one-loop RG computation is not enough to reproduce the quark wave function renormalization, although the one-loop quark mass function exhibits a qualitative agreement with lattice results. The higher loop corrections or other ignored effects are thus highly significant to describe the quark sector well. Here, we try to examine the quark propagator within the framework of the one-loop RG to catch the qualitative feature as a first attempt.

 \begin{figure}[tb]
 \begin{minipage}{\hsize}
  \begin{center}
   \includegraphics[width=\linewidth]{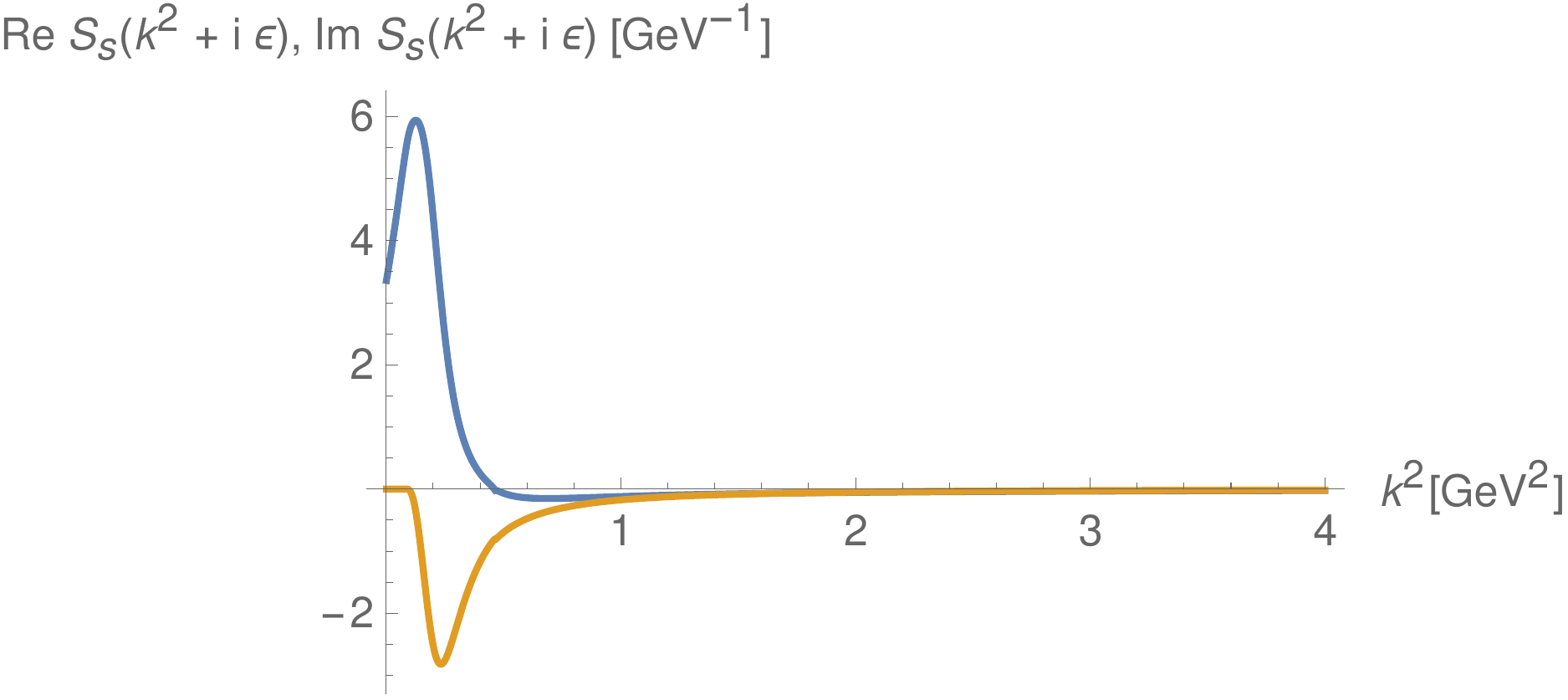}
  \end{center}

 \end{minipage}
 \begin{minipage}{\hsize}
  \begin{center}
   \includegraphics[width=\linewidth]{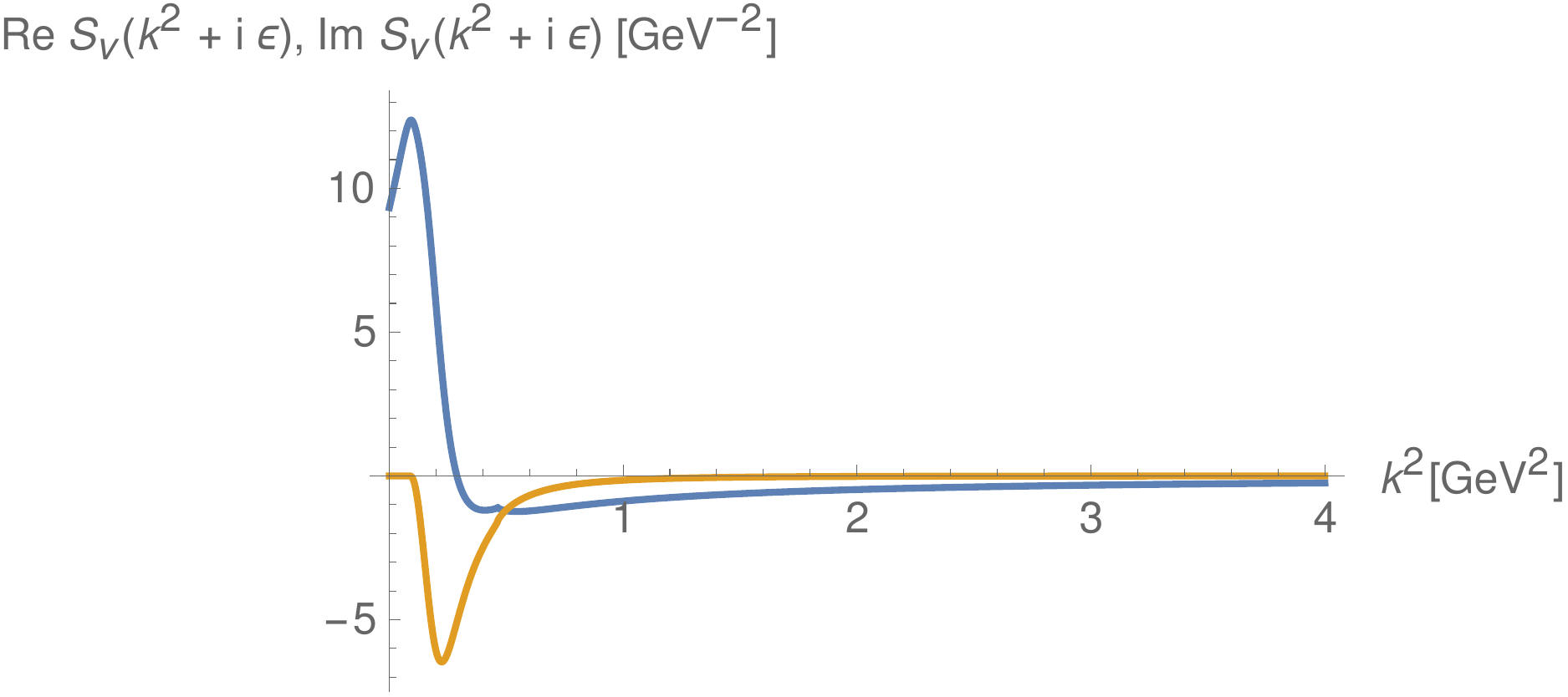}
  \end{center}
 \end{minipage}
 \caption{Real (blue) and imaginary (orange) parts of the scalar part (top) and the vector part (bottom) of the quark propagator on the positive real axis at the realistic parameters for $N_F = 2$ \cite{PTW14}. Their spectral function are negative, from which $N_W(C) = -2$ for both the scalar and vector parts.}
    \label{fig:realistic-quark}
\end{figure}

 \begin{figure}[tb]
  \begin{center}
   \includegraphics[width=\linewidth]{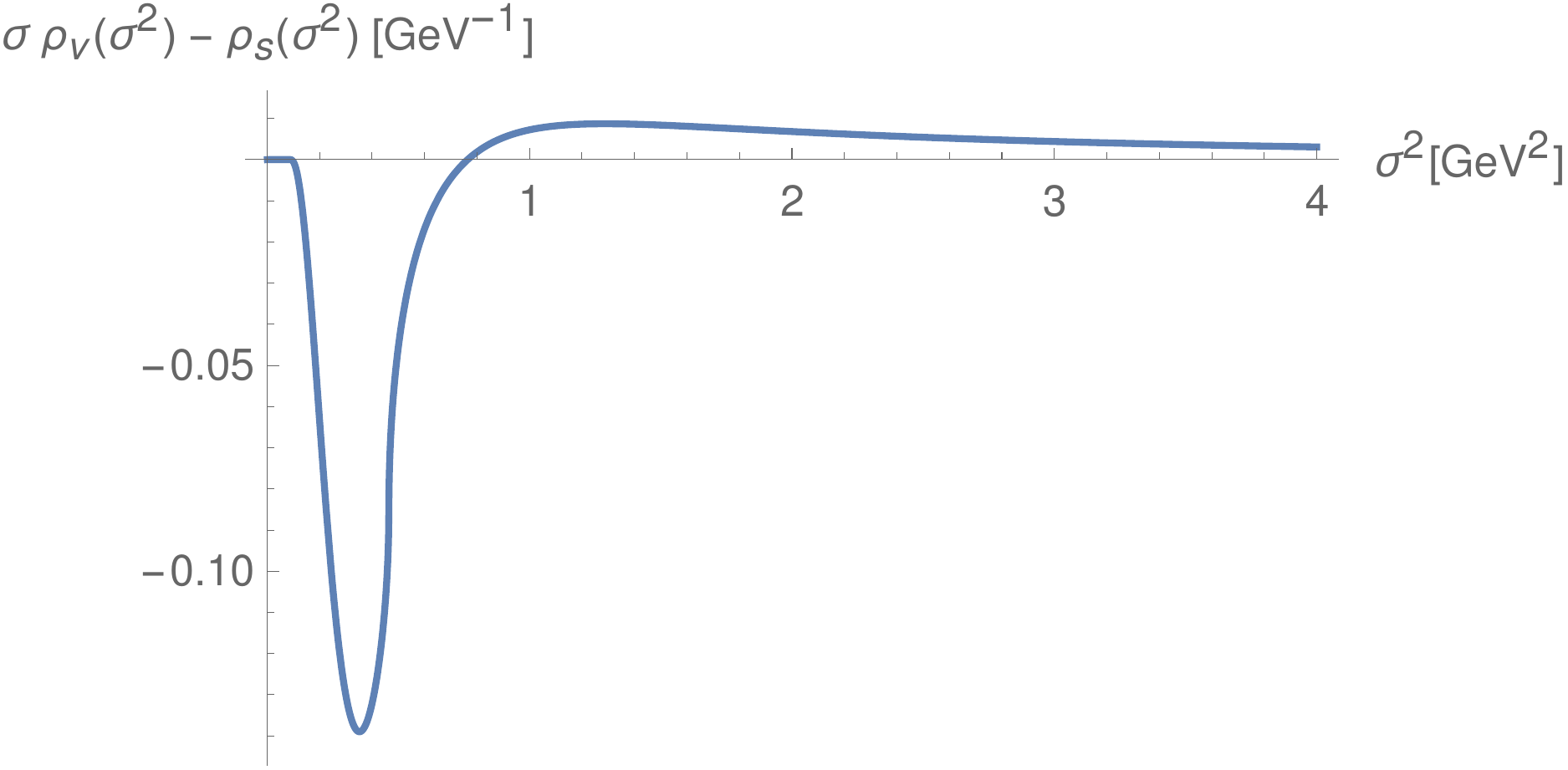}
  \end{center}
 \caption{The quark spectral function $\sigma \rho_v(\sigma^2) - \rho_s(\sigma^2)$ is plotted at the realistic parameters. The positivity condition is violated below $0.8$ GeV$^2$. }
    \label{fig:realistic-quark-positivity-2}
\end{figure}

Let us consider the vector and scalar parts of the quark propagator:
\begin{align}
\mathcal{S}_v(k_E^2) = \frac{\Gamma_{v}^{ren.}}{k_E^2 (\Gamma_{v}^{ren.})^2 + (\Gamma_{s}^{ren.})^2}, \notag \\
\mathcal{S}_s(k_E^2) = \frac{\Gamma_{s}^{ren.}}{k_E^2 (\Gamma_{v}^{ren.})^2 + (\Gamma_{s}^{ren.})^2}. \label{eq:quark-prop-def}
\end{align}
We define the spectral functions associated to these propagators,
\begin{align}
\rho_v(\sigma^2) &= \frac{1}{\pi} \operatorname{Im} \mathcal{S}_v(k_E^2 = - \sigma^2 - i \epsilon), \notag \\ \rho_s(\sigma^2) &= \frac{1}{\pi} \operatorname{Im} \mathcal{S}_s(k_E^2 = - \sigma^2 - i \epsilon).
\end{align}
Note that the positivity of the state space would imply
\begin{align}
\rho_v(\sigma^2) > 0, ~~ \sigma \rho_v(\sigma^2) - \rho_s(\sigma^2) > 0. \label{eq:positivity-quark-prop}
\end{align}

We use the same improvement scheme as (\ref{eq:RG_improvement_scheme}). The scalar and vector parts of the quark propagator at the realistic values of the parameters are plotted in Fig. \ref{fig:realistic-quark}. Both of the propagators have negative spectral functions and therefore have one pair of complex conjugate poles like the gluon propagator. Notice that both the conditions for the positivity (\ref{eq:positivity-quark-prop}) are violated. The former condition can be seen from the vector part of the quark propagator in Fig.~\ref{fig:realistic-quark}. Figure \ref{fig:realistic-quark-positivity-2} shows the violation of the latter one.

 \begin{figure}[tb]
 \begin{minipage}{\hsize}
  \begin{center}
   \includegraphics[width=0.8\linewidth]{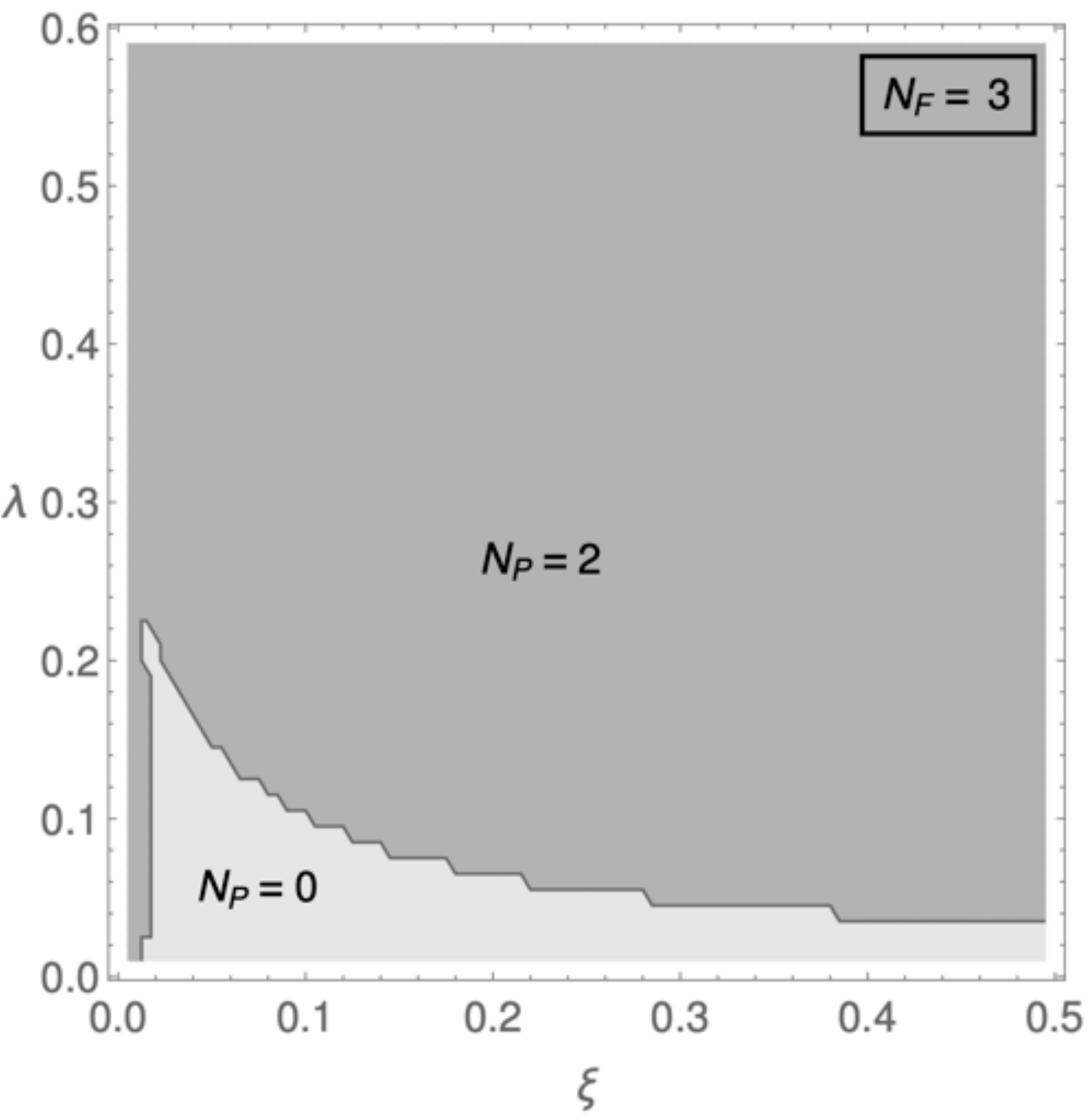}
  \end{center}

 \end{minipage}
 \begin{minipage}{\hsize}
  \begin{center}
   \includegraphics[width=0.8\linewidth]{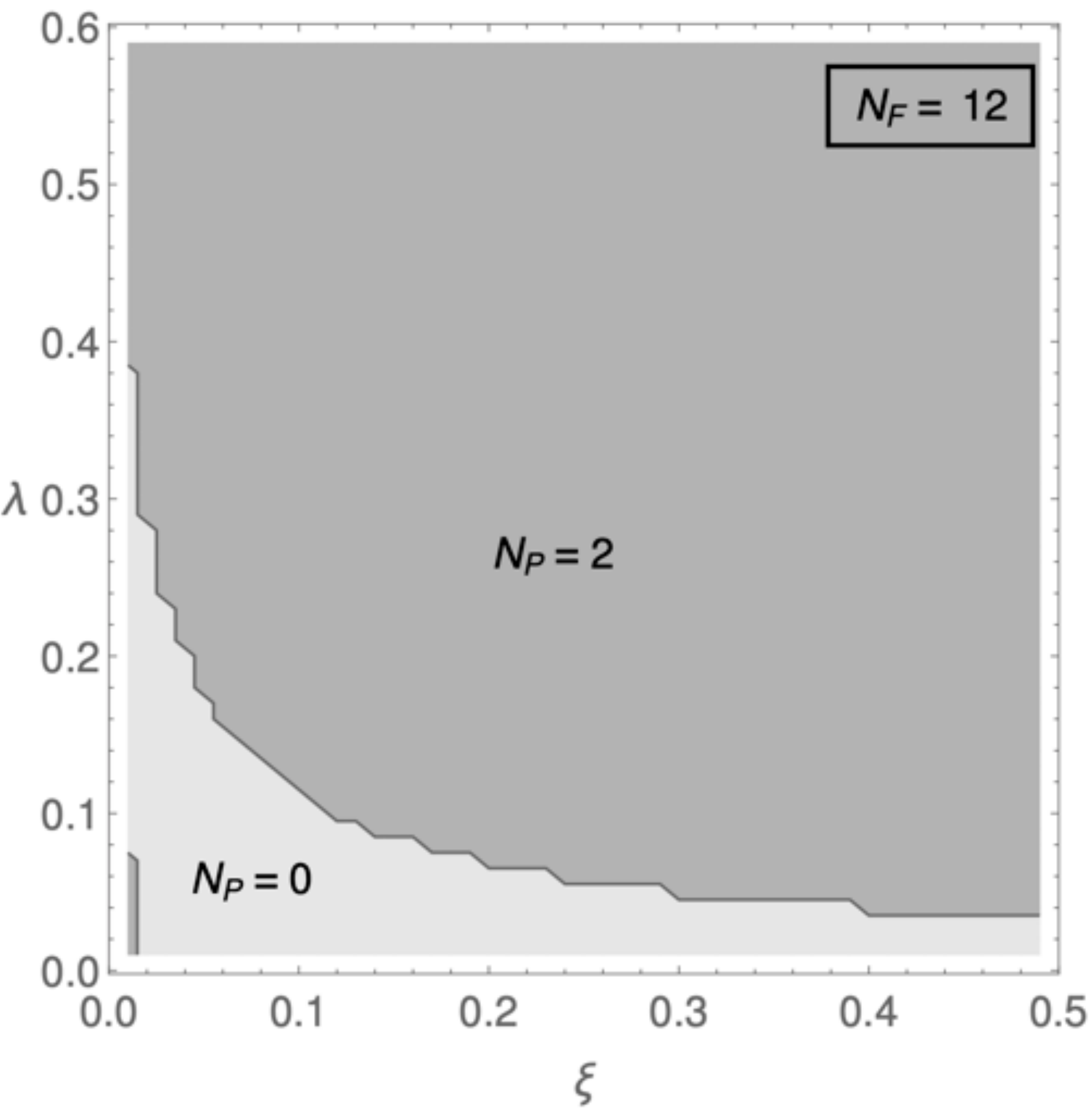}
  \end{center}
 \end{minipage}
 \caption{Two-dimensional slice of equi-$N_W(C)$ volume of the quark propagator at $N_F = 3$ (top) and $N_F = 12$ (bottom) in the $(\lambda = \frac{C_2(G) g^2}{16 \pi^2}, M^2/\mu_0^2 = 0.2, \xi= \frac{m_q^2}{M^2})$ space. Both of the scalar and vector parts provide the same $N_W(C)$.}
    \label{fig:RG-one-loop-nf312}
\end{figure}

Next, we investigate the number of complex poles with various parameters $(N_F,g,M,m_q)$. The results on the winding number $N_W(C)$ at $N_F = 3$ and $N_F = 12$ are shown in Fig. \ref{fig:RG-one-loop-nf312}. We numerically checked that both of the scalar and vector parts of the quark propagator (\ref{eq:quark-prop-def}) yield the same $N_W(C)$ on the region shown in Fig. \ref{fig:RG-one-loop-nf312}, from which $N_Z = 0$ for the both parts.

The quark sector does not exhibit a new structure by adding $N_F$ quarks. The qualitative insensitivity of the quark sector to $N_F$ is evident in the sense that the strict one-loop expression is, of course, independent of $N_F$ and that the quark propagator will only be influenced indirectly by $N_F$ via the running parameters. However, it is worthwhile to note that the region with $N_P = 0$ extends slightly as $N_F$ increases, in particular for very light quarks $\xi \ll 1$, which is a common feature with the gluon one. 

\subsection{Ghost propagator}
Finally, let us add some comments on the ghost propagator. The strict one-loop propagator is not affected by the dynamical quarks. Therefore, from the proposition (Case III) of \cite{HK2018}, the value $N_W(C) = 0$ for ghosts will hold after the RG improvements unless the RG trajectory has a Landau pole. Therefore, we conclude that the analytic structure of the ghost propagator within this approximation is insensitive to $N_F$ and that the ghost propagator has no complex poles.

\section{Conclusion}

Let us summarize our findings.
The argument principle for a propagator relates the propagator on the timelike momenta and the number of complex poles. The winding number $N_W(C) = N_Z - N_P$ can be computed numerically from a propagator on timelike momenta according to (\ref{eq:winding-timelike}), where $N_Z$ and $N_P$ stand for the number of complex zeros and poles respectively.

To study the analytic structures of the QCD propagators, we have employed an effective model of QCD, the massive Yang-Mills model. We have found that the infrared safe trajectories, on which the running coupling is finite in all scales, reproduce the UV asymptotic behavior (\ref{eq:UV_asymptotic}) originally obtained by Oehme and Zimmermann \cite{OZ80} and that the gluon spectral function is negative in the IR limit $\rho (\sigma^2) \sim - \sigma^2$ for $m_q > 0$ for quarks with the quark mass parameter $m_q$ of this model (\ref{eq:infrared_negativity_gluon}). The UV negativity of Oehme-Zimmermann and IR negativity of the gluon propagator argued in this paper supports the existence of complex poles in the gluon propagator from the general relationship claiming that a negative spectral function in a weak sense leads to complex poles~\cite{HK2018}.

In the ``realistic'' parameters used in \cite{PTW14} to fit the numerical lattice results, both the gluon and quark propagators have quasinegative spectral functions and one pair of complex conjugate poles, while the ghost propagator has no complex poles.

Finally, we have investigated the number of complex poles in this model with many quarks by computing $N_W(C)$ for various parameters. For the gluon, there are several interesting features. First, the $N_P = 2$ region dominates the parameter region if $\xi = m_q^2/M^2$ is not so small or $N_F \lesssim 4$. In this region, the gluon propagator has one pair of complex conjugate poles as in the pure Yang-Mills case \cite{HK2018}.
Second, the $N_P = 4$ region, where the gluon propagator has two pairs of complex conjugate poles, expands by adding light quarks. A typical set of the parameters ($g \approx 4, M^2 \approx 0.2$ GeV$^2$) is covered by the $N_P = 4$ region when $4 \lesssim N_F \leq 9, 0.2 \lesssim \xi \lesssim 0.6$. These features hold in the computations both for the strict one-loop and the RG-improved one-loop results. Third, the strict one-loop result on the locations of complex poles indicates that the gluon tends to be ``particlelike'' as $N_F$ increases. Fourth, for $N_F \geq 10$, the one-loop RG-improved gluon propagator has no $N_P = 4$ region. Moreover, the gluon propagator with $N_F~(N_F \geq 10)$ very light quarks $\xi \ll 1$ has no complex poles even for a typical value of the gauge coupling.

The existence of complex poles invalidates the K\"all\'en-Lehmann spectral representation, which is a fundamental consequence of QFT describing physical particles. Therefore, the existence of complex poles of the gluon and quark propagators could be a signal of confinement of the elementary degrees of freedom in QCD.

In conclusion, the above results imply that the confinement mechanism may depend on the number of quarks and quark mass since complex poles represent a deviation from physical particles and will be related to confinement. Furthermore, the drastic change between $N_F = 9$ and $N_F = 10$ quarks could be possibly related to the ``deconfinement'' in line with the conformal window.

\section{Discussion and future work}

Several comments regarding these results are in order.

First, let us mention a comparison with a similar approach \cite{Siringo16a,Siringo16b}, where the analytic structures of gluon and quark propagators are investigated with a massive-type model in light of the variational principle and optimization. In there, the gluon propagator has two pairs of complex conjugate poles, while the quark propagator has a timelike pole and no complex poles in $N_F = 2$ QCD. These structures are different from ours, in which the gluon and quark propagators have one pair at the ``realistic'' parameter. Although the quark sector of both models lacks accuracy, the difference will be relevant in light of the confinement of quark degrees of freedom\footnote{The confinement of quark degrees of freedom, which stands for absence of the quark one-particle state from the physical spectrum, should not be confused with the well-studied ``quark confinement'' that means an external quark source requires infinite energy or the linear rising quark-antiquark potential.}; a timelike pole might correspond to a physical one-particle state even after some confinement mechanism works. In this sense, the absence of a timelike pole will be favored.

Second, we comment on the $N_F$ dependence of the condensation.
We have studied the mass-deformed model as an effective model for QCD or QCD-like theories based on the facts that the gluon mass can minimally improve the gauge fixing and that the effective potential for the operator $\mathscr{A}_\mu \mathscr{A}_\mu$ calculated by the local composite operator technique indicates the condensation of this operator \cite{Verschelde2001,BG2003}. The former argument can give a mass-like effect independently upon $N_F$. However, the latter one will be substantially affected by the presence of quarks. The effective potential of \cite{BG2003} appears to be ``unbounded'' for $\gamma_0 > 0$, or $N_F \geq 10$ for $G = SU(3)$. 
The ``unboundedness'' follows from the fact that the coefficient of the Hubbard-Stratonovich transformation $\zeta$ runs into the negative infinity $\zeta \rightarrow - \infty$ in the UV limit $\mu \rightarrow \infty$ due to additive counterterms for $N_F \geq 10$ even if it is set to be a positive value at some scale. This problem might indicate the limitation of the perturbative treatment for the local composite operator. Therefore, we have no cogent argument supporting the dimension-two gluon condensate for $N_F \geq 10$.

Third, related to the second remark, the validity of our results will be questionable for large $N_F$. For instance, two-loop corrections will be important if the first coefficient of the beta function is small as in the original argument of the infrared conformality \cite{Caswell,BZ82}. Moreover, the results from the truncated Schwinger-Dyson equation \cite{conformal-window-sd} show that the gluon and ghost propagators seem to obey a scaling-type power law in a wide range of momentum in the conformal window; the description by the massive Yang-Mills model will be inappropriate above the critical value of $N_F$. However, we can expect that the massive Yang-Mills model will be valid in the QCD-like phase, or below the critical value of $N_F$. Therefore, the massive Yang-Mills model may capture some information on the transition from the QCD-like phase.

Fourth, $N_F = 10$, where the first coefficient of the gluon anomalous dimension changes its sign, is the value at which the analytic structure of the gluon propagator changes drastically in our analysis. This value appears in various perspectives. For example, there has been some proposal that the negativity of the gluon anomalous dimension is crucial for confinement \cite{Nishijima}. $N_F \approx 10$ can be the critical value of the conformal phase transition \cite{conformal-window-frg}.

Finally, the investigation of the analytic structures by model calculations is speculative and should be taken as an attempt toward capturing some aspects of the intricate dynamics of QCD. Many works of literature have different claims. For example, reference \cite{ADFM2004} claims that the quark propagator has a pole at a timelike momentum while the gluon propagator has complex poles by using some parameterizations for the propagators and the numerical solution of truncated Dyson-Schwinger equation. The gluon propagator obtained by solving the truncated Dyson-Schwinger equation on the complex momentum plane in pure Yang-Mills theory is shown to have no complex poles \cite{SFK12}. A recent reconstruction technique indicates complex poles in the gluon propagator \cite{BT2019}.
Pursuing the rich analytic structure of the QCD propagators deserves further investigations because a QFT describing confined particles is not yet well understood.

On a formal side, a local QFT cannot yield complex poles in the standard perspective, see e.g., \cite{JLDrepr}. One might assert that complex poles correspond to short-lived excitations and break the locality and unitarity in the level of propagators \cite{Stingl96,HKRSW}. However, if we analytically continue\footnote{To our knowledge, a method to reconstruct a QFT from a given Euclidean field theory has not been established in the presence of complex poles. The standard reconstruction does not work due to the violation of the reflection positivity. \cite{AQFT,KSOMH18}} a propagator with complex poles not in the complex momentum but in the complex time from the Euclidean space to the Minkowski one, the resulted propagator can be interpreted as a propagator of a QFT with an indefinite metric state space having complex spectra that can satisfy local commutativity, e.g., \cite{Nakanishi71}. Notice that such theories with complex spectra and an indefinite metric are out of the scope of the axiomatic quantum field theory because their Wightman functions are not tempered distribution due to complex energies, and the theorems derived by assuming the temperedness are not applicable to the ``propagators with complex poles''. Then, complex poles would not lead to the nonlocality and just represent unphysical degrees of freedom. Further discussion on this issue is reserved for future works. As discussed in \cite{Zwanziger90,BDGHSVZ10}, it would also be interesting to study how the complex poles are ``canceled'' in the physical propagator.

\section*{Acknowledgements}
This work was supported by Grant-in-Aid for Scientific Research, JSPS KAKENHI Grant Number
(C) No.~19K03840.
Y. H. is supported by JSPS Research Fellowship for Young Scientists.

\appendix

\section{Counting complex poles for various infrared behaviors}

Here, we generalize the IR asymptotic condition (ii) $D(k^2 = 0) > 0$ to derive the formula (\ref{eq:winding-timelike}) in Sec.~II. This will be relevant for the scaling solution of the gluon propagator or a propagator having massless singularity.

Incidentally, in regard to the IR behavior of the gluon propagator, there is an argument on the gluon propagator in the Landau gauge \cite{Zwanziger2013}:
If one restricts the configuration space inside the first Gribov region, the gluon propagator satisfies $\lim_{k \rightarrow +0} k^{d-2} D(k) = 0$, where $d$ is the spacetime dimension. This excludes the massless free behavior for the gluon propagator irrespective of $N_F$ for $d \leq 4$. Note that, however, this proposition is shown for theories avoiding the Gribov ambiguity by the restriction; this proposition is not applicable to theories that fix the gauge by schemes averaging over Gribov copies like \cite{ST12} or some justification of the standard Faddeev-Popov Lagrangian.

The generalization of (\ref{eq:winding-timelike}) is as follows.

Suppose that the propagator $D(k^2)$ and its data $\{ D(x_n + i \epsilon) \}_{n=1}^N$ satisfies the following three conditions:
\begin{enumerate}
 \item In the UV limit $|z| \rightarrow \infty$, $D(k^2)$ has the same phase as the free propagator, i.e., $\arg (-D(z)) \rightarrow  \arg \frac{1}{z} $ as $|z| \rightarrow \infty$.
 \renewcommand{\labelenumi}{(\roman{enumi}')} 
 
 \item $D(k^2) \rightarrow Z_{IR} (-k^2)^\alpha$ as $|k^2| \rightarrow 0$, where $\alpha$ is a real number.
 \renewcommand{\labelenumi}{(\roman{enumi})}
 
 \item $\{k^2 = x_n + i \epsilon \}_{n=0}^N$ is sufficiently dense so that $D(k^2= x + i \epsilon)$ changes its phase at most half-winding ($\pm \pi$) between $x_n+i\epsilon$ and $x_{n+1}+i\epsilon$, where we denote sufficiently small $x_0 = \delta^2 >0$ and sufficiently large $x_{N+1} = \Lambda^2$, on which we will take the limits $\delta^2 \rightarrow +0$ and $\Lambda^2 \rightarrow +\infty$.
\end{enumerate}

Then, the winding number $N_W(C)$ is expressed as
\begin{align}
N_W (C) = - \alpha -1  + 2 \sum_{n=0}^N \frac{1}{2\pi} \operatorname{Arg}\left[ \frac{D(x_{n+1}+i\epsilon)}{D(x_n + i \epsilon)}\right]. \label{eq:winding-timelike-appendix}
\end{align}
Let us derive this expression. First, we decompose the path around the positive real axis $C_2$ into three pieces $C_2 = C_{2,+} \cup C_\delta \cup C_{2,-}$, where $C_{2,\pm}$ stands for the path along the positive real axis of $C_{2,\pm} = \{x \pm i \epsilon ; \delta^2 < x < \Lambda^2 \}$ and $C_\delta$ for the small circle whose center is the origin $k^2 = 0$.
Accordingly, the winding number can be decomposed into the integrals
\begin{align}
N_W (C) = N_W(C_\delta) + N_W(C_1) + N_W(C_{2,-}) + N_W(C_{2,+}).
\end{align}
The integral $ N_W(C_1) + N_W(C_{2,-}) + N_W(C_{2,+})$ can be evaluated as before,
\begin{align}
 N_W(C_1) &+ N_W(C_{2,-}) + N_W(C_{2,+}) \notag \\
 &= -1  + 2 \sum_{n=0}^N \frac{1}{2\pi} \operatorname{Arg}\left[ \frac{D(x_{n+1}+i\epsilon)}{D(x_n + i \epsilon)}\right].
\end{align}
For the contribution from the small circle, note that $\frac{D(k^2 \pm i \epsilon)}{|D(k^2 \pm i \epsilon)|} = e^{\mp i \pi \alpha}$ as $k^2 \rightarrow +0$. Therefore the phase factor $D/|D|$ varies from $e^{+i \pi \alpha }$ to  $e^{-i \pi \alpha }$, from which
\begin{align}
N_W(C_\delta) = - \alpha.
\end{align}

To sum up, we obtain (\ref{eq:winding-timelike-appendix}). Note that the IR suppression contributes negatively to $N_W(C) = N_Z - N_P$.

\section{Detailed results on the one-loop gluon propagator for various parameters}

Here, we show detailed analyses on the results given in Sec. IV A 3 on the strict one-loop gluon propagator. To check the insensitivity of the transition, we shall compute $N_W(C)$ in the three dimensional parameter space $(\lambda = \frac{C_2(G) g^2}{16 \pi^2}, u = \frac{M^2}{\mu_0^2}, \xi= \frac{m_q^2}{M^2})$.

\begin{figure}[tb]
 \begin{minipage}{\hsize}
  \begin{center}
   \includegraphics[width=0.8\linewidth]{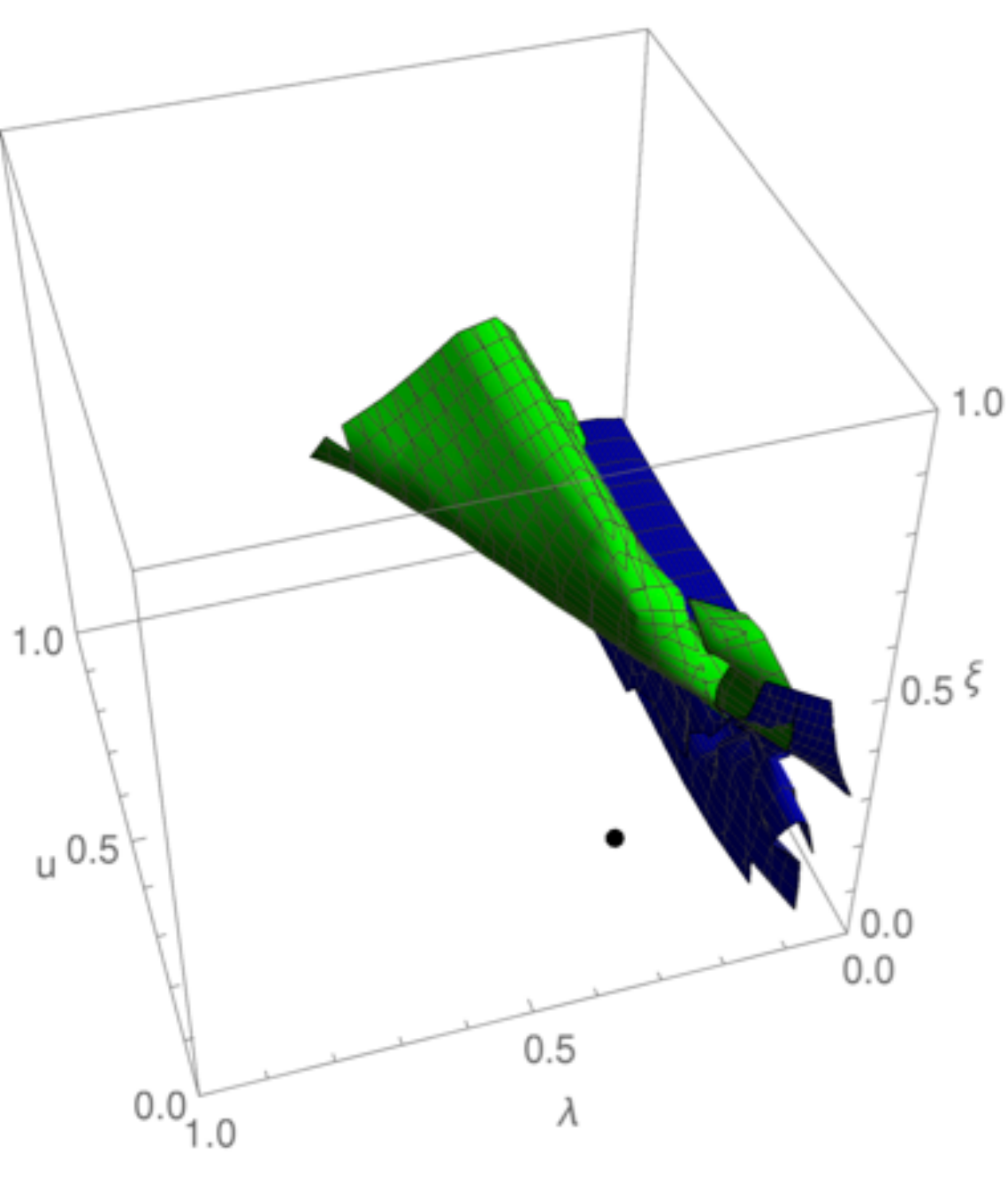}
  \end{center}

 \end{minipage}
 \begin{minipage}{\hsize}
  \begin{center}
   \includegraphics[width=0.8\linewidth]{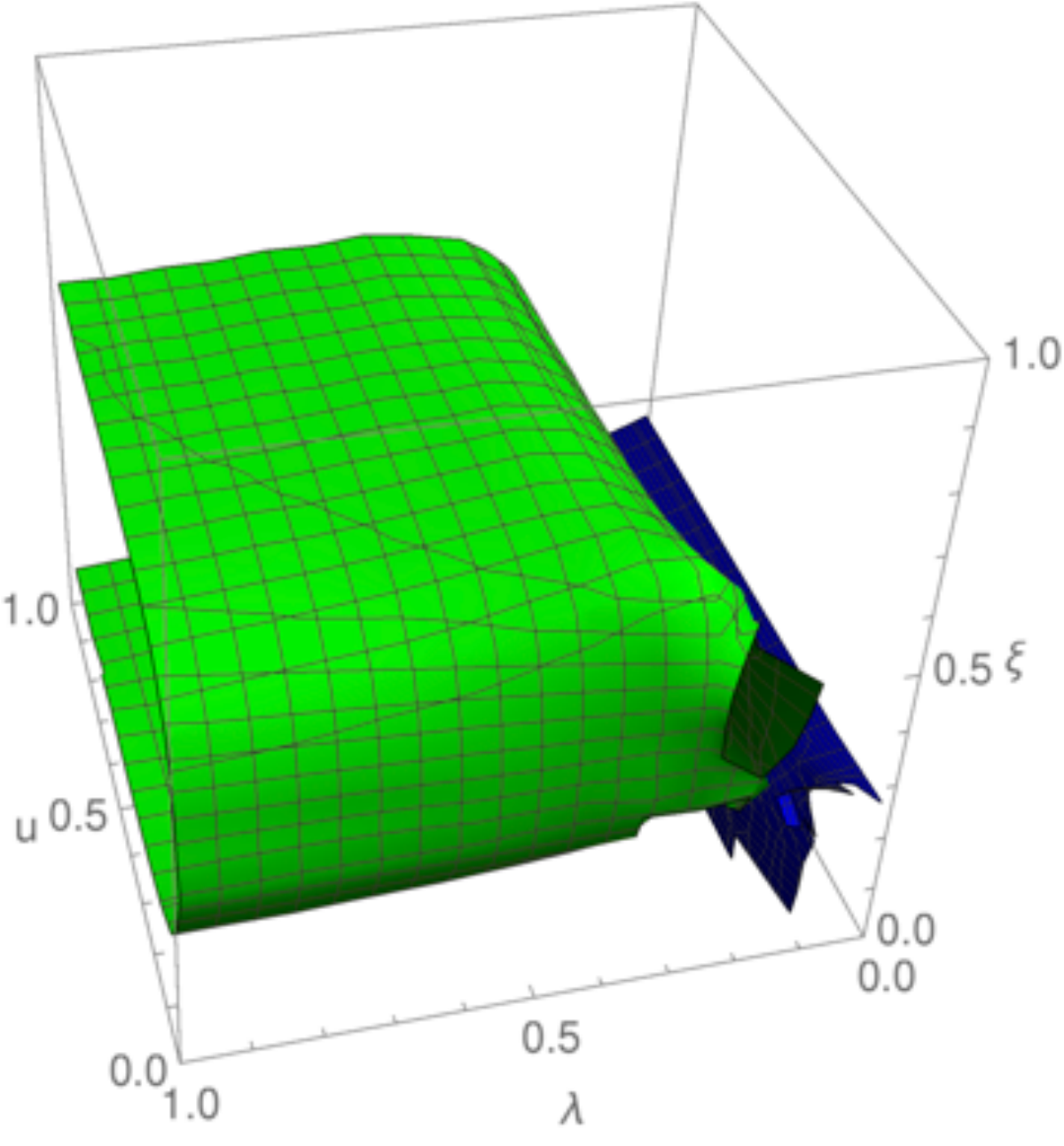}
  \end{center}
 \end{minipage}
 \caption{Boundary surfaces of equi-$N_W(C)$ volume at $N_F = 3$ (top) and $N_F = 6$ (bottom) in the $(\lambda = \frac{C_2(G) g^2}{16 \pi^2}, u = \frac{M^2}{\mu_0^2}, \xi= \frac{m_q^2}{M^2})$ space. Their cross sections at several values of $\xi$ are shown in Fig.~\ref{fig:nf3mgdep-fixedxi} and Fig.~\ref{fig:nf6mgdep-fixedxi}.
 The gluon propagator has four complex poles inside the green surface and no complex poles inside the blue surface; otherwise the gluon propagator has two complex poles. At $N_F = 3$, the black dot shows a typical set of the parameters ($g = 4$, $M^2 = 0.2~ \mathrm{GeV}^2$, and $\xi = 0.1$) and the gluon propagator has two complex poles around there. On the other hand, at $N_F = 6$, the gluon propagator has $N_P = 4$ for $0.2 \lesssim \frac{m_q^2}{M^2} \lesssim 0.6$ around the typical values $g \sim 4$, $M^2 \sim 0.2~ \mathrm{GeV}^2$.
 }
    \label{fig:mgdependence}
\end{figure}

Figure \ref{fig:mgdependence} plots the boundary surfaces of equi-$N_W(C)$ volume at $N_F = 3$ and $N_F = 6$ in the three-dimensional parameter space $(\lambda, u, \xi)$. The gluon propagator has four complex poles inside the green surface ($0.2 \lesssim \xi \lesssim 0.6$) and no complex poles inside the blue surface (with small $\lambda$ and $\xi$), and two complex poles except these regions.

\begin{figure}[t]
 \begin{minipage}{\hsize}
  \begin{center}
   \includegraphics[width=0.8\linewidth]{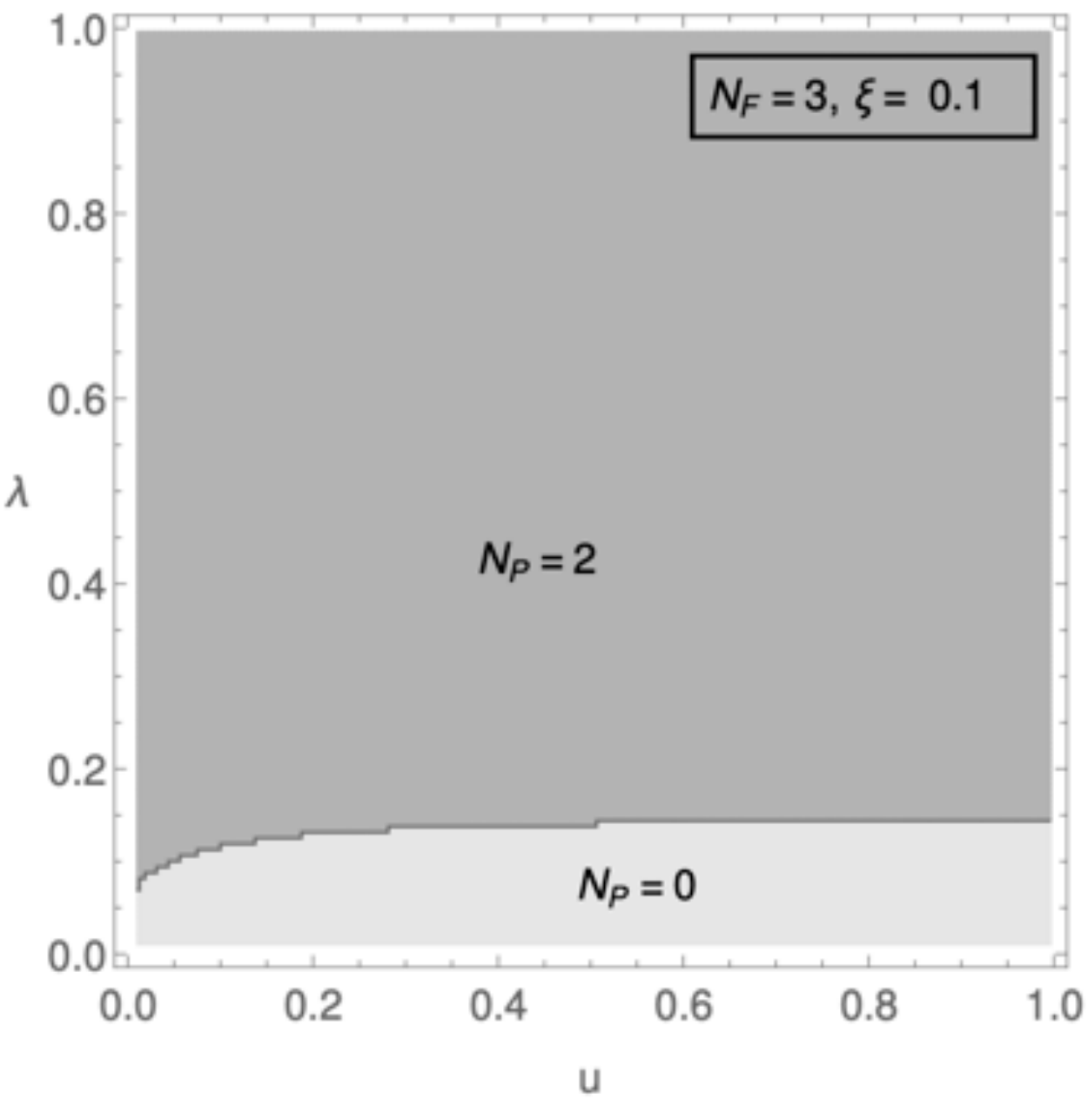}
  \end{center}
 \end{minipage}
 
  \begin{minipage}{\hsize}
  \begin{center}
   \includegraphics[width=0.8\linewidth]{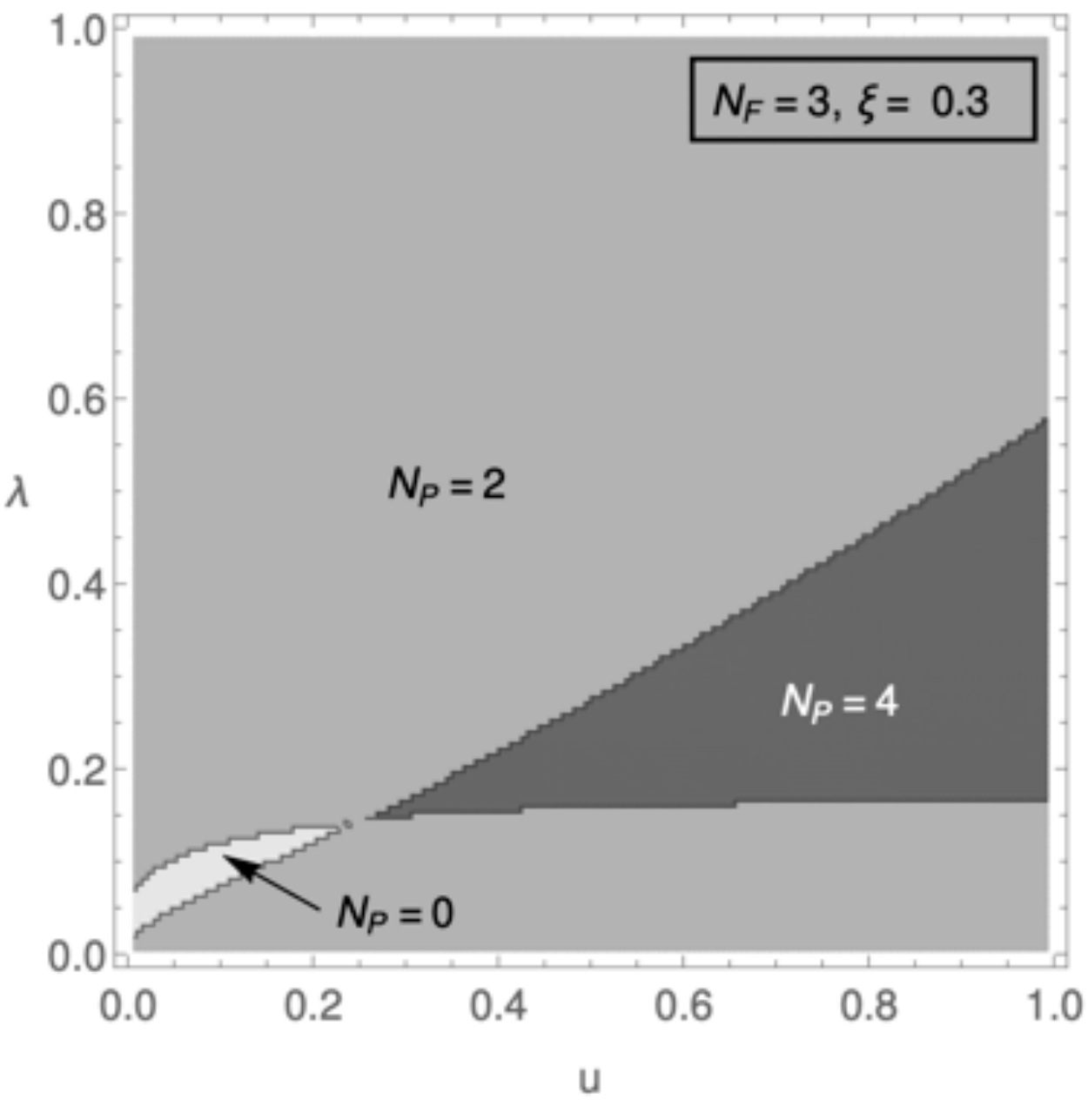}
  \end{center}
 \end{minipage}
 
  \begin{minipage}{\hsize}
  \begin{center}
   \includegraphics[width=0.8\linewidth]{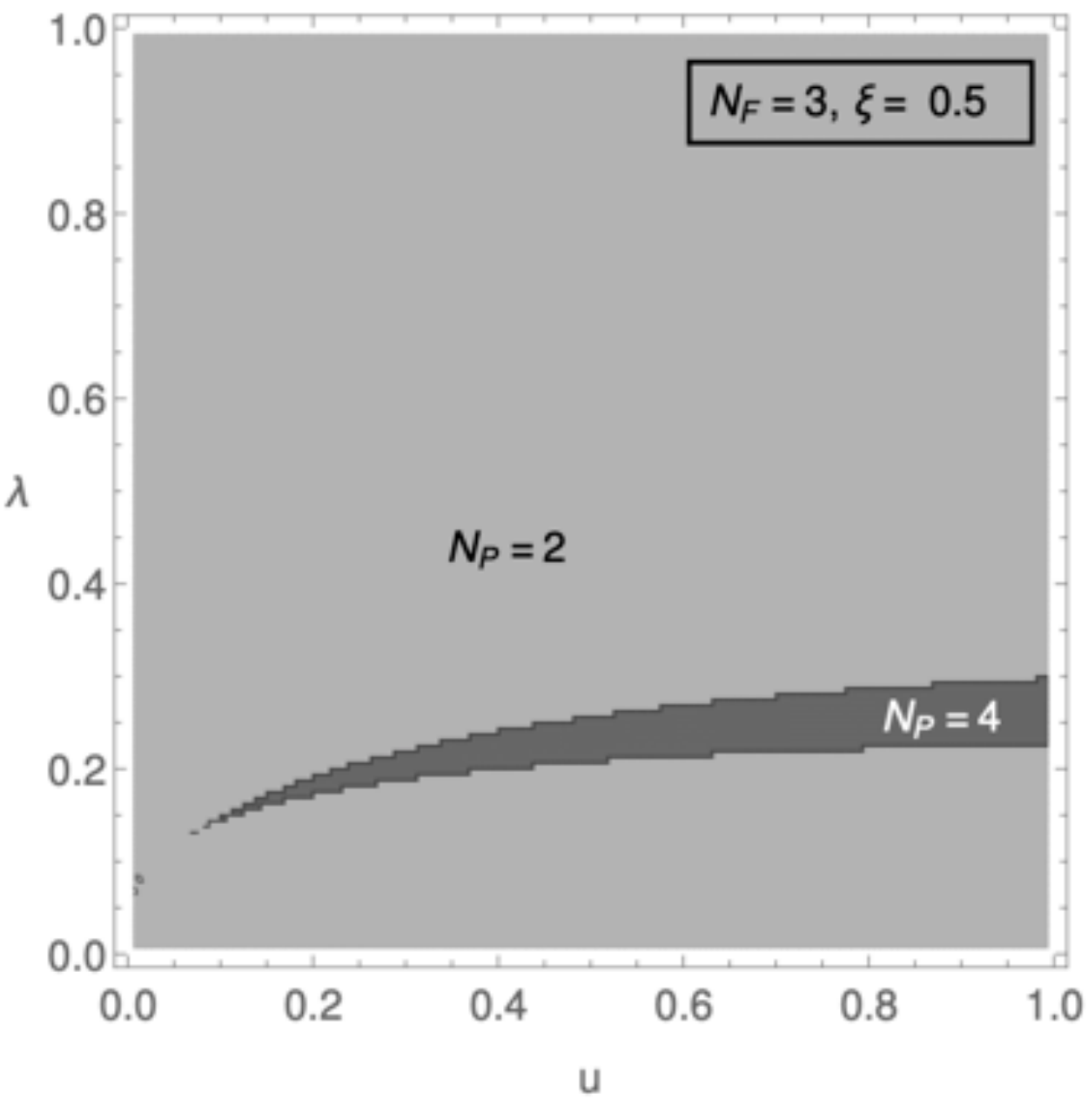}
  \end{center}
 \end{minipage}

 \caption{Contour plots of $N_W(C) = -N_P$ of the gluon propagator on the two-dimensional parameter space $(\lambda = \frac{C_2(G) g^2}{16 \pi^2}, u = \frac{M^2}{\mu_0^2})$ at $\xi = 0.1,0.3,0.5$ for $N_F = 3$.
 }
    \label{fig:nf3mgdep-fixedxi}
\end{figure}

\begin{figure}[t]
 \begin{minipage}{\hsize}
  \begin{center}
   \includegraphics[width=0.8\linewidth]{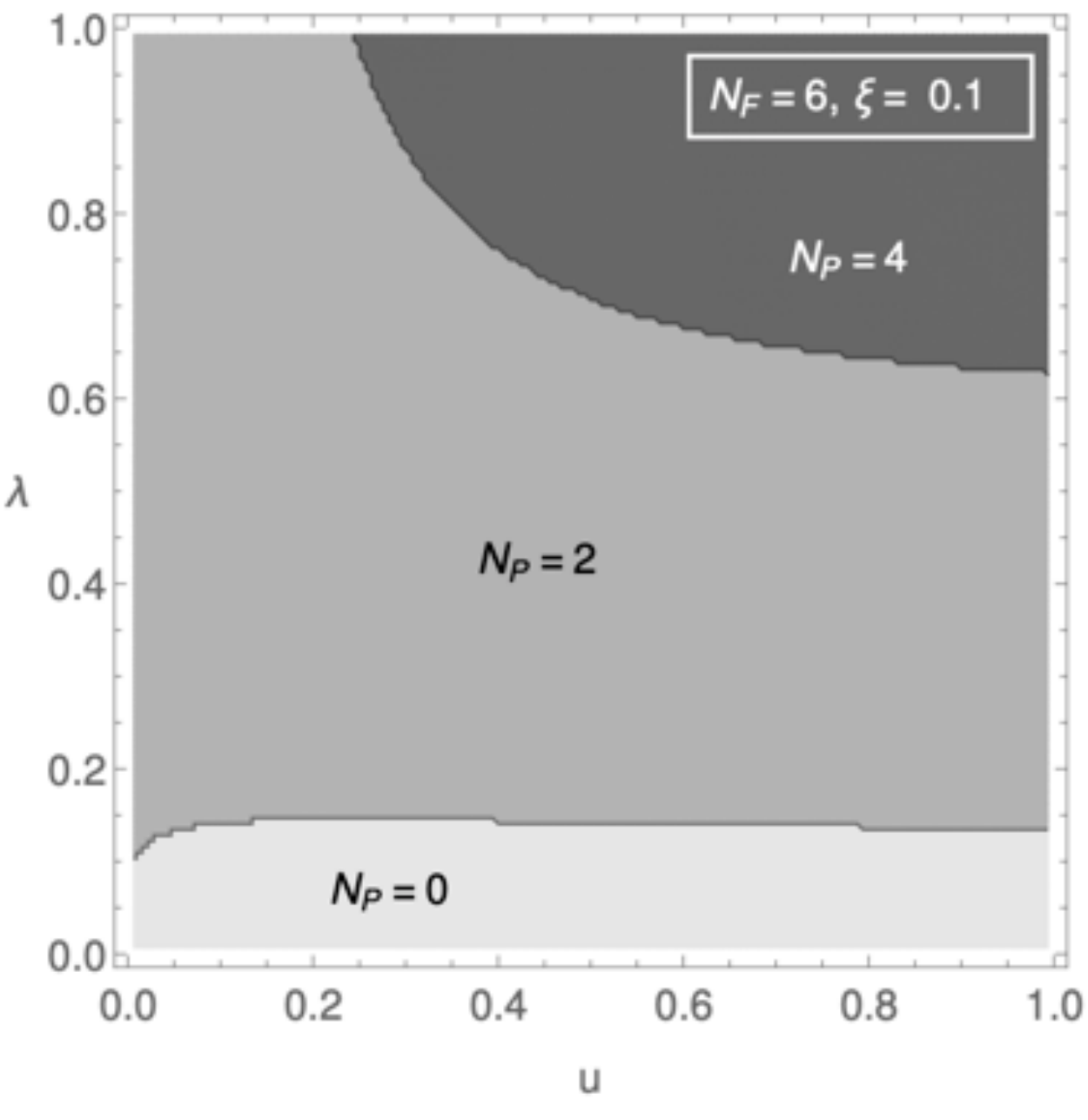}
  \end{center}
 \end{minipage}
 
  \begin{minipage}{\hsize}
  \begin{center}
   \includegraphics[width=0.8\linewidth]{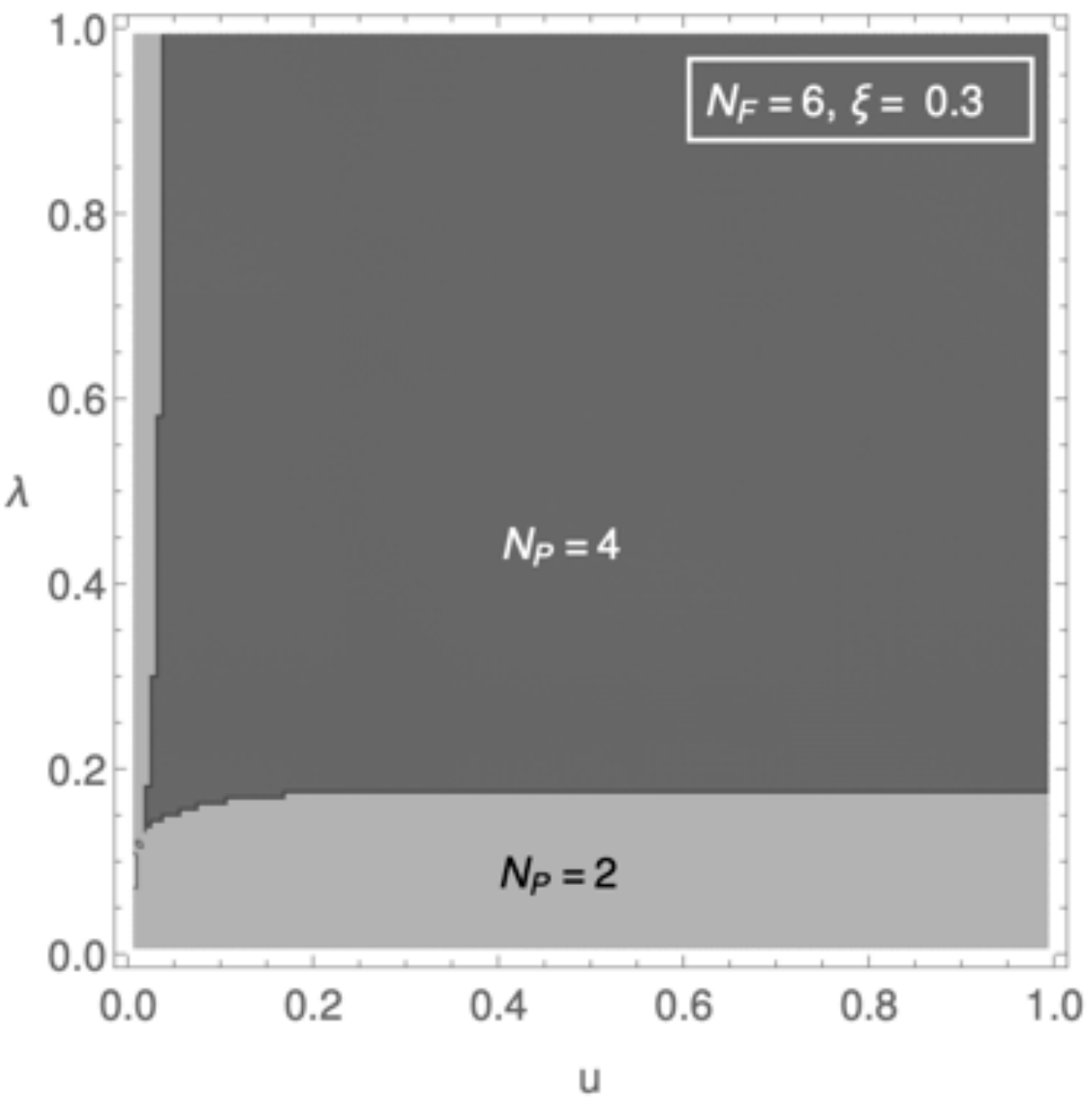}
  \end{center}
 \end{minipage}
 
  \begin{minipage}{\hsize}
  \begin{center}
   \includegraphics[width=0.8\linewidth]{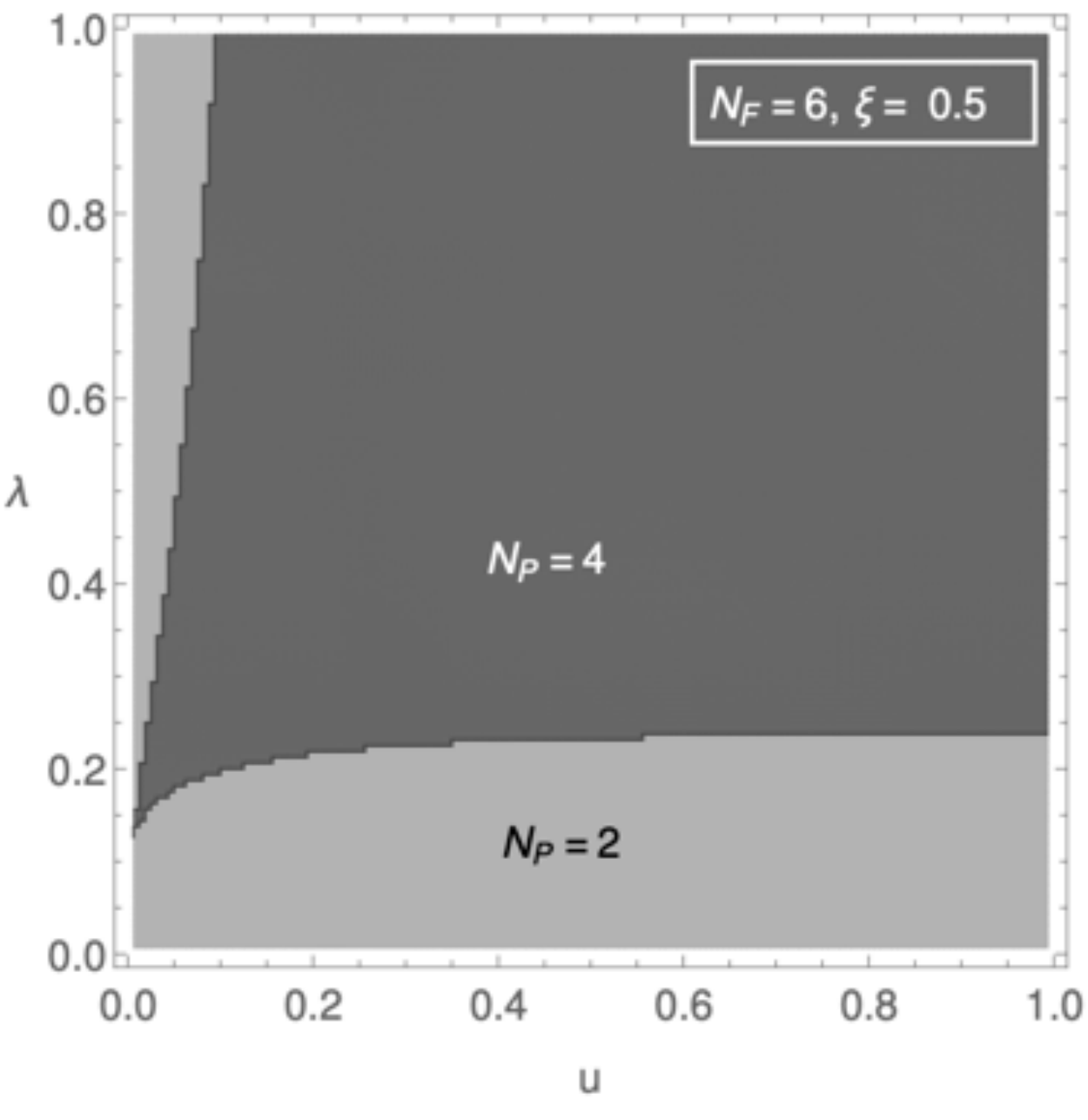}
  \end{center}
 \end{minipage}

 \caption{Contour plots of $N_W(C) = -N_P$ of the gluon propagator on the two-dimensional parameter space $(\lambda = \frac{C_2(G) g^2}{16 \pi^2}, u = \frac{M^2}{\mu_0^2})$ at $\xi = 0.1,0.3,0.5$ for $N_F = 6$.
 }
    \label{fig:nf6mgdep-fixedxi}
\end{figure}

Contour plots for selected $\xi$ are displayed in Fig.~\ref{fig:nf3mgdep-fixedxi} for $N_F =3$ and in Fig.~\ref{fig:nf6mgdep-fixedxi} for $N_F = 6$. They give two-dimensional slices at fixed $\xi = 0.1,0.3,0.5$ of Fig.~\ref{fig:mgdependence}. The regions colored with light gray ($N_P = 0$) in Fig.~\ref{fig:nf3mgdep-fixedxi} and Fig.~\ref{fig:nf6mgdep-fixedxi} are two-dimensional cross sections at fixed $\xi$ of the volumes inside the blue surfaces in the top and bottom figures of Fig.~\ref{fig:mgdependence}, respectively. Similarly, the regions colored with dark gray ($N_P = 4$) in Fig.~\ref{fig:nf3mgdep-fixedxi} and Fig.~\ref{fig:nf6mgdep-fixedxi} are those of the volumes surrounded by the green surfaces in the top and bottom figures of Fig.~\ref{fig:mgdependence}, respectively.

Figure~\ref{fig:mgdependence} demonstrates the $N_W(C) = -4$ region inside the green surface expands from $N_F = 3$ to $N_F = 6$. In the figure of $N_F = 3$ (top of Fig.~\ref{fig:mgdependence}), the set of the typical values ($g = 4$, $M^2 = 0.2~ \mathrm{GeV}^2$, and $\xi = 0.1$) is shown as a black dot, of which Fig.~\ref{fig:flavor_quasiposineg_windingnumber} is computed at $(g, M^2)$.
The $N_W(C) = -4$ region occupies the region $0.2 \lesssim \frac{m_q^2}{M^2} \lesssim 0.6$ around the typical values of the parameters ($g \sim 4$, $M^2 \sim 0.2~ \mathrm{GeV}^2$) at $N_F = 6$.
From these observations, we deduce that the existence of the transition is insensitive to a detailed choice of $g$ and $M^2$. Therefore, the rough investigation in Sec.~IV A will be qualitatively valid in this model.

In addition, since the gluon propagator acquires a pole at timelike momentum on the boundary of the $N_P = 4$ region and its poles will move like Fig.~\ref{fig:flavor_pole_location_B},  $w/v$ for the pole of the gluon propagator decreases as $N_F$ increases. Hence, the gluon will become more particlelike in the presence of many light quarks, irrespectively of a detailed choice of the parameter $(g,M)$.

\end{document}